\documentclass[12pt]{iopart}
\usepackage[dvipdfmx,hiresbb]{graphicx}
\usepackage{amsmath}
\usepackage{amssymb}
\usepackage{amscd}
\usepackage{color}

\usepackage[mathscr]{eucal}

\newtheorem{theorem}{Theorem}
\newtheorem{lemma}[theorem]{Lemma}

\newtheorem{proposition}[theorem]{Proposition}

\newtheorem{definition}[theorem]{Definition}
\newtheorem{example}[theorem]{Example}

\newtheorem{remark}[theorem]{Remark}

\newcommand{\Z}{{\mathbb Z}}
\newcommand{\R}{{\mathbb R}}

\newcommand{\Rig}{{\rm Rig}}

\newcommand{\wt}{{\rm wt}}

\newcommand{\proof}{\noindent \textit{Proof. }}
\newcommand{\qed}{\hfill $\Box$}
\newcommand{\Trees}{{\boldsymbol T}{\boldsymbol r}{\boldsymbol e}{\boldsymbol e}}
\newcommand{\Root}{{\rm Root}}
\newcommand{\Tree}{{\rm Tree}}
\newcommand{\MTree}{{\rm MTree}}
\newcommand{\oRoot}{\overline{\rm Root}}
\newcommand{\Sub}{{\rm Sub}}
\newcommand{\MSub}{{\rm MSub}}
\newcommand{\oSub}{\overline{\rm Sub}}
\newcommand{\oSubo}{\overline{\rm Sub}_{\rm o}}
\newcommand{\oSube}{\overline{\rm Sub}_{\rm e}}
\newcommand{\oRooto}{\overline{\rm Root}_{\rm o}}
\newcommand{\oRoote}{\overline{\rm Root}_{\rm e}}

\begin{document}

\title{Combinatorial aspects of the conserved quantities of the tropical periodic Toda lattice}

\author{Taichiro Takagi}
\address{Department of Applied Physics, National Defense Academy, Kanagawa 239-8686, Japan}
%
%
\begin{abstract}
The tropical periodic Toda lattice (trop p-Toda) is a dynamical system attracting attentions in the area of the interplay of integrable systems and tropical geometry.
We show that the 
Young diagrams associated with
trop p-Toda
given by two very different definitions are identical.
The first definition is given by a Lax representation of
the discrete periodic Toda lattice, and the second one is
associated with a generalization of the Kerov-Kirillov-Reshetikhin bijection in the combinatorics of the Bethe ansatz.
By means of this identification, it is shown for the first time that the Young diagrams given by the latter definition are preserved
under time evolution.
This result is regarded as an important first step in clarifying the iso-level set structure of this dynamical system in {general} cases, 
i.~e.~not restricted to {generic} cases. 
\end{abstract}


\section{Introduction}\label{sec:1}
The Toda lattice is one of the most famous integrable systems in classical mechanics \cite{Toda}.
Recently, one of its variations is attracting attentions
in the context of connections between tropical geometry and integrable systems \cite{AMS12}.
We call this system the tropical periodic Toda lattice (trop p-Toda) \cite{IKT12, T2012, II12}.
Its evolution equation was known as the ultra-discretization of the discrete periodic Toda equation \cite{KT02}.
In \cite{IT08, IT09}, Inoue and Takenawa studied this system and clarified its iso-level set structure under a certain condition
which they call {\em generic}.
From the viewpoint of tropical geometry,
this condition is related to the smoothness of the tropical spectral curve determined by the conserved quantities of the system.

In this paper we study conserved quantities of trop p-Toda 
{without the generic condition}.
In particular, 
we show that the 
Young diagrams associated with
trop p-Toda
given by two very different definitions are identical.
From one of the definitions one immediately sees that the common Young diagram is preserved under time evolution.
In the context of integrable cellular automaton explained below, this Young diagram represents the content of solitons
in the system, and
the generic condition is requiring no two solitons have a common amplitude. 
We believe that this identification of the Young diagrams
is the first step to clarify the iso-level set structure of this dynamical system in {\em general} cases, 
i.~e.~not restricted to {\em generic} cases.

The two definitions of the Young diagrams are as follows.
The first  is related to Lax representation of the  discrete periodic Toda lattice (dp-Toda).
Here the $k$th conserved quantity of dp-Toda
is defined as a sum of products of $k$ dependent variables whose indices obey a
nearest-neighbor-exclusion condition. 
Then the  corresponding conserved quantity of trop p-Toda
is defined as its tropical limit or
{\em tropicalization} \cite{BK00,IMS2009,Mac12}.
We show that the above condition
leads to a condition of weak convexity relating the conserved quantities,
which is also a new result of this paper that
enables us to represent them as a Young diagram.
The second is related to a generalization of the Kerov-Kirillov-Reshetikhin (KKR) bijection in combinatorics of Bethe ansats \cite{KKR, KR},
especially one of its variations in $\mathfrak{sl}_2$ case \cite{Takagi1}.
It is also considered as a continuous analogue of the 
`10-elimination' algorithm for conserved quantities of an integrable cellular automaton known as the periodic box-ball system 
(pBBS) \cite{YYT}.

We note that for a special case there has already been attention paid to this remarkable equivalence
of Young diagrams.
The pBBS is regarded as a case of trop p-Toda, where the values of its dependent variables are restricted to positive integers.
In this case, the equivalence of  the Young diagrams has been pointed out by Iwao and Tokihiro \cite{IT07}.
On the basis of their idea of drawing diagrams associated with the second definition of the Young diagram,
we give a proof of our main theorem
for the conserved quantities of
general trop p-Toda.

Here we explain
the reason why we expect that the identification of the Young diagrams from
Lax representation and those from generalized KKR bijection
is the first step to clarify the iso-level set structure of trop p-Toda without the generic condition.
In the pBBS case, the KKR bijection gives the action-angle variables of this dynamical system \cite{KTT}.
While the action variables are the conserved quantities, the angle variables yield certain time evolutions which turn out to be flows on the iso-level set.
By this fact the author has succeeded to clarify the iso-level set structure of pBBS without the generic condition \cite{T2010}.
Since the trop p-Toda is a generalization of the pBBS, it is reasonable to consider the corresponding generalization of the KKR bijection.
We note that in the pBBS case
the conservation of the Young diagrams defined by KKR bijection is directly proved by using the crystal theory, combinatorial $R$ maps, and Yang-Baxter relations (See Theorem 2.2 and Proposition 3.4 of \cite{KTT}).
Since these methods are not developed in trop p-Toda case, our main result of this paper is so far the only proof of the conservation of the Young diagrams defined by the generalized KKR. 

Readers may wonder why trop p-Toda is worth studying independently 
of the already known many results in pBBS.
The most remarkable difference 
between pBBS and trop p-Toda
is that
the iso-level set of the latter is not a finite set but is an algebraic variety. 
This implies that
while
any state comes back to the same state
in pBBS, that is not true in trop p-Toda.
Actually, when the lengths of the solitons are linearly independent over the field of rational numbers,
the phase flow can be dense in the iso-level set
as in the case of classical mechanics \cite{Abook}.
However,
this is nothing but only one aspect of the fact that
their iso-level sets are totally different mathematical
objects.
A really important problem here is that
the structure of the iso-level set of trop p-Toda has not yet been fully clarified in general. 
As we have mentioned, it has been clarified only in the case when 
it reduces to a real torus under
the generic condition.
Without this condition, we have no suitable description of
their connected components or invariant tori,
which have different sizes according to their internal symmetries.
We expect that they should not be regarded as mere subsets as in the pBBS case \cite{T2010}, but should be
regarded as lower dimensional tori embedded in the whole
iso-level set. 
The present work is a starting point to developing such a description,
which will contribute to
making progress in the studies on
tropical geometry and integrable systems.

This paper is organized as follows.
In \S \ref{sec:2_1} we derive the nearest neighbor exclusion condition on the indices of variables from a determinant formula of a matrix.
In \S \ref{sec:2_2} definitions of dp-Toda and trop p-Toda are given, and
we show that the matrix given above is related to the Lax matrix of dp-Toda.
Here we obtain conserved quantities of dp-Toda and trop p-Toda by the formulae described by the nearest neighbor exclusion condition. 
In \S \ref{sec:2_3} we show that 
the conserved quantities of trop p-Toda
are in weak convexity condition
(Theorem \ref{th:jun10_1}), which
enables us to describe the conserved quantities as a Young diagram.
In \S \ref{sec:2_4} another algorithm to construct the Young diagram is introduced, and
the main result of this paper (Theorem \ref{th:main}) is presented. 
We devote our efforts to proving this theorem in \S \ref{sec:3}.
In \S \ref{sec:3_1} we introduce an algorithm to draw diagrams of trees that visualizes the algorithm in \S \ref{sec:2_4}.
In \S \ref{sec:3_2} some elementary lemmas on the properties of the diagrams are presented.
Using these lemmas, we give a proof of the main theorem in \S \ref{sec:3_3}, leaving proofs of two more lemmas which we call Close Packing Lemma and Forest Realization Lemma.
We devote our efforts to proving these lemmas in \S \ref{sec:3_4} and \S \ref{sec:3_5}.
A continuous analogue of the KKR bijection
is discussed in \S \ref{sec:4}.
Some concluding remarks are given in \S \ref{sec:5}.
\section{Discrete and tropical periodic Toda lattice}
\label{sec:2}
\subsection{Determinant formulas and the nearest neighbor exclusion condition}
\label{sec:2_1}
Throughout this paper we use the symbol
$\triangleleft$
by the following meaning:
\begin{equation}
i \triangleleft j \Leftrightarrow i+1<j.
\end{equation}
Given a sequence of real numbers $a_0(=0), a_1, a_2, \ldots$, let $c_k^{(N)}$ be the numbers defined by the recursion relation
\begin{equation}\label{eq:nov11_1}
c_k^{(N)} = c_k^{(N-1)} +(a_{2N-1}+a_{2N})c_{k-1}^{(N-1)}
-a_{2N-2} a_{2N-1} c_{k-2}^{(N-2)},
\end{equation}
and the boundary conditions
\begin{equation}\label{eq:nov11_2}
c_k^{(N)}=0 \quad \mbox{for} \,  k<0 \quad \mbox{or}\quad k>N,
\qquad
c_0^{(N)} = 1 \quad \mbox{for}\, N \geq 0.
\end{equation}
Then it is easy to see that the unique
solution of \eqref{eq:nov11_1} under
\eqref{eq:nov11_2} is given by
\begin{equation}\label{eq:nov11_3}
c_k^{(N)} = \sum_{1 \leq i_1 \triangleleft \, i_2 \triangleleft \dots \triangleleft \, i_k \leq 2N}
a_{i_1} a_{i_2}\cdots a_{i_k}.
\end{equation}
Let $e_k^{(N)}$ be the numbers defined by the relations
\begin{align}\label{eq:nov11_4a}
&e_1^{(N)} = c_1^{(N)}, \qquad 
e_2^{(N)} = c_2^{(N)}- a_1 a_{2N},\\
\label{eq:nov11_4b}
&e_k^{(N)} = c_k^{(N)} 
- a_1 a_{2N}
\sum_{3 \leq i_1 \triangleleft \, i_2 \triangleleft \dots \triangleleft \, i_{k-2} \leq 2N-2}
a_{i_1} a_{i_2}\dots a_{i_{k-2}},
\end{align}
for $3 \leq k \leq N$.
Then it is easy to see that
\begin{equation}\label{eq:nov11_5}
e_k^{(N)} = \sum_{\stackrel{1 \leq i_1 \triangleleft \, i_2 \triangleleft \dots \triangleleft \, i_k \leq 2N}{\scriptstyle (i_1,i_k) \ne (1,2N)}}
a_{i_1} a_{i_2}\dots a_{i_k}.
\end{equation}

Let $F_1(x;a_1,a_2)=x+a_1+a_2$ and
\begin{equation}\label{eq:july1_1}
F_N(x;a_1,\dots,a_{2N}) =
\det
\begin{pmatrix}
x+a_1+a_2 & 1 & & & \\
a_2 a_3 & x+a_3+a_4 & 1 & & \\
 & a_4 a_5 & \ddots &\ddots & \\
 & & \ddots & \ddots & 1 \\
 & & & a_{2N-2} a_{2N-1} & x + a_{2N-1} + a_{2N}
\end{pmatrix},
\end{equation}
for $N \geq 2$.
\begin{lemma}[\cite{Tbook}, Proposition 7.1]\label{lem:july2_6}
\begin{equation}
F_N(x;a_1,\dots,a_{2N}) = \sum_{k=0}^N c_k^{(N)} x^{N-k}.
\end{equation}
\end{lemma}
\proof
Let $F_N(x;a_1,\dots,a_{2N}) = \sum_{k=0}^N b_k^{(N)} x^{N-k}$.
Then we have $b_0^{(N)} = 1$.
By expanding \eqref{eq:july1_1} with respect to its $N$th row one obtains
\begin{eqnarray}\label{eq:july2_1}
F_N(x;a_1,\dots,a_{2N}) &=& (x + a_{2N-1}+a_{2N}) F_{N-1}(x;a_1,\dots,a_{2N-2}) \nonumber \\
&-& a_{2N-2} a_{2N-1} F_{N-2}(x;a_1,\dots,a_{2N-4}).
\end{eqnarray}
Defining $b_k^{(N)}=0 \, \mbox{for} \,  k<0 \, \mbox{or}\, k>N$, one can deduce from \eqref{eq:july2_1} that
the $b_k^{(N)}$'s satisfy the same recursion relation \eqref{eq:nov11_1} as $c_k^{(N)}$'s
for $1 \leq k \leq N$.
Hence $b_k^{(N)} = c_k^{(N)}$.
\qed

Let
\begin{equation}\label{eq:july2_3}
G_N(x;a_1,\dots,a_{2N}) =
\det
\begin{pmatrix}
x+a_1+a_2 & 1 & & & (-1)^{N-1}a_1 a_{2N}/y \\
a_2 a_3 & x+a_3+a_4 & 1 & & \\
 & a_4 a_5 & \ddots & \ddots & \\
 & & \ddots & \ddots & 1 \\
(-1)^{N-1}y & & & a_{2N-2} a_{2N-1} & x + a_{2N-1} + a_{2N}
\end{pmatrix},
\end{equation}
for $N \geq 3$.
\begin{lemma}
\begin{equation}
G_N(x;a_1,\dots,a_{2N}) =
y+ (\prod_{i=1}^{2N} a_i)/y +
\sum_{k=0}^N e_k^{(N)} x^{N-k}.
\end{equation}
\end{lemma}
\proof
By expanding \eqref{eq:july2_3} with respect to its $N$th column one obtains
\begin{eqnarray}\label{eq:july2_4}
G_N(x;a_1,\dots,a_{2N}) &=& (x + a_{2N-1}+a_{2N}) F_{N-1}(x;a_1,\dots,a_{2N-2}) \nonumber \\
&&+ y -  a_{2N-2} a_{2N-1} F_{N-2}(x;a_1,\dots,a_{2N-4}) \nonumber \\
&&+ (a_1 \dots a_{2N})/y - a_1 a_{2N} F_{N-2}(x;a_3,\dots,a_{2N-2}) \nonumber \\
&=& F_N(x;a_1,\dots,a_{2N})- a_1 a_{2N} F_{N-2}(x;a_3,\dots,a_{2N-2}) \nonumber \\
&&+ y + (a_1 \dots a_{2N})/y,
\end{eqnarray}
where we have used \eqref{eq:july2_1}.
Let $G_N(x;a_1,\dots,a_{2N}) =
y+ (\prod_{i=1}^{2N} a_i)/y +
\sum_{k=0}^N g_k^{(N)} x^{N-K}$.
From Lemma \ref{lem:july2_6} and 
\eqref{eq:july2_4}, one finds that
the $g_k^{(N)}$'s satisfy the same relations, \eqref{eq:nov11_4a} and \eqref{eq:nov11_4b}, as the $e_k^{(N)}$'s
for $1 \leq k \leq N$.
Hence $g_k^{(N)} = e_k^{(N)}$.
\qed
\subsection{The evolution equation and Lax representation}
\label{sec:2_2}
On the basis of \cite{IKT12},
we briefly review the derivation of discrete and tropical periodic Toda lattice equations.
Let $\{ x_n(t) \}_{n \in \Z_N}$ be a set of smooth functions of time $t \in \R$.
Set $a_{2n} = a_{2n}(t) := 1 + \delta \dot{x}_n(t),
a_{2n+1} = a_{2n+1}(t) := \delta^2 e^{x_{n+1}(t)-x_n(t)}$
with $\delta > 0$ and $\bar{a}_j = a_j(t + \delta)$ for $j \in \Z_{2N}$.
Then we have $\lim_{\delta \rightarrow 0} \frac{1}{\delta^3} (\bar{a}_{2n} \bar{a}_{2n+1} - a_{2n+1} a_{2n+2})=0$.
Suppose $x_n = x_n(t)$'s satisfy the Toda lattice equation
\begin{equation}
\ddot{x}_n = e^{x_{n+1}-x_n} - e^{x_{n}-x_{n-1}}.
\end{equation}
Then we have $\lim_{\delta \rightarrow 0} \frac{1}{\delta^2}(\bar{a}_{2n-1}+\bar{a}_{2n} - a_{2n}-a_{2n+1})=0$.
Under this consideration,
we define the evolution equations for the
 {\em discrete periodic Toda lattice} as
\begin{equation}\label{eq:jun30_1}
\bar{a}_{2n-1}+\bar{a}_{2n} = a_{2n}+a_{2n+1}, \quad
\bar{a}_{2n} \bar{a}_{2n+1} = a_{2n+1} a_{2n+2},
\end{equation}
where the $a_n = a_n^t, \bar{a}_n = a_n^{t+1}$ are dependent variables which depend on discrete spatial coordinate $n \in \Z_{2N}$ and discrete time $t \in \Z$.
Obviously, $\prod_{l=1}^{2N} a_l$ is a conserved quantity.
By lemma \ref{le:nov11_6} we will find that
$h_N = \prod_{l=1}^{N} a_{2l-1} + \prod_{l=1}^{N} a_{2l}$ is also a conserved quantity.
This implies that
$(\prod_{l=1}^{N} \bar{a}_{2l-1}, \prod_{l=1}^{N} \bar{a}_{2l})=(\prod_{l=1}^{N} a_{2l},\prod_{l=1}^{N} a_{2l-1})$ or $(\prod_{l=1}^{N} a_{2l-1},\prod_{l=1}^{N} a_{2l})$.
While the former leads to the trivial solution $\bar{a}_n = a_{n+1}$, 
the latter to a non-trivial  solution
\begin{equation}\label{eq:jun30_2}
\begin{split}
&\bar{a}_{2n} = a_{2n+1} + a_{2n}
\frac{1-\prod_{l=1}^N (a_{2l-1}/a_{2l})}{\sum_{k=0}^{N-1} \prod_{l=1}^k (a_{2(n-l)+1}/a_{2(n-l)}) } ,\\
&\bar{a}_{2n+1} = a_{2n+1} a_{2n+2}/\bar{a}_{2n}.
\end{split}
\end{equation}
For a derivation of this solution, see Proposition 6.13 of \cite{IKT12}.
Now we consider its tropicalization, which is a procedure to replace $\times$ by $+$, and $+$ by $\min$.
Note that the numerator in \eqref{eq:jun30_2} is a conserved quantity.
By regarding it as a positive constant and setting it to be zero
under the {\em tropicalization with
trivial valuation} \cite{Mac12},
we obtain a dynamical system given by the 
piecewise linear evolution equations
\begin{equation}\label{eq:july2_7}
\begin{split}
&\bar{A}_{2n} = \min \left( A_{2n+1}, A_{2n} - 
\min_{0 \leq k \leq N-1} \left( \sum_{l=1}^k (A_{2(n-l)+1} - A_{2(n-l)} ) \right) \right),\\
&\bar{A}_{2n+1} = A_{2n+1} + A_{2n+2} - \bar{A}_{2n},
\end{split}
\end{equation}
on the phase space
$\displaystyle{
\mathcal{T} 
= 
\left\{(A_n)_{n \in \Z_{2N}} ~\Bigg|~ \sum_{l=1}^N A_{2l} < \sum_{l=1}^N A_{2l-1} \right\}
\subset \R^{2N}.}
$
We call this system {\em the tropical periodic Toda lattice} \cite{IKT12}.
\begin{remark}
We have changed the notations as
$a_{2n-2} = q_n, a_{2n-1}=w_n,A_{2n-2} = Q_n, A_{2n-1}=W_n$ from those in \cite{IKT12}, since this enables us to
describe the conserved quantities neatly.
\end{remark}

\begin{figure}[hbtp]
\centering
\scalebox{0.5}[0.5]{
\includegraphics[clip]{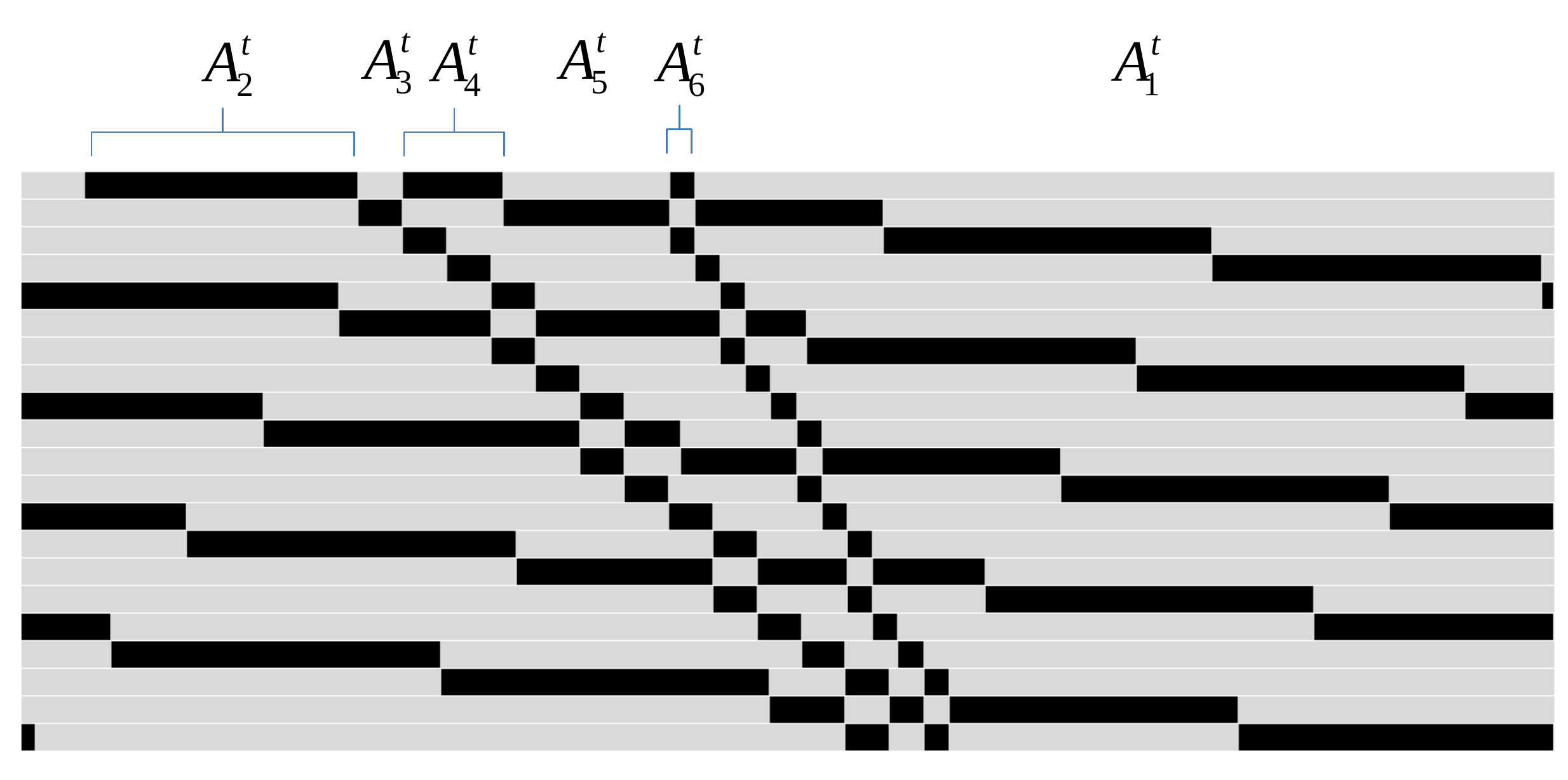}}
\caption{A representation for the time evolution of trop p-Toda. 
Let $t$ be the time for the top row, where 
the lengths of three black segments (`solitons') are $A_2^t, A_4^t, A_6^t$,
and the lengths of white segments between them 
(with regard to the periodic boundary) are $A_3^t, A_5^t, A_1^t$.
In the second row
the length of the first black segment to the right of
the $j$-th `soliton' at time $t$ is  $A_{2j}^{t+1}$. 
In the same way, the strip at the $n$-th row visualizes the quantities   $A_j^{t+n-1}$.}
\label{fig:0}
\end{figure}

Without loss of generality,
we can assume 
all the $A$-variables in
\eqref{eq:july2_7} take their values in $\mathbb{R}_{>0}$.
This enables us to represent the time evolution of trop p-Toda by
a sequence of two-colored (white and black) strips,
where the lengths of the white (resp.~black) segments are denoted by
$A_{2n-1}$s (resp.~$A_{2n}$s).
See Figure \ref{fig:0} for an example.



Now on the basis of \cite{IT08} we briefly review the
Lax representation for the discrete periodic Toda lattice equation.
Let
\begin{equation}
R(\lambda) = 
\begin{pmatrix}
1 &   & & & a_1 /\lambda \\
 a_3 & 1 &   & & \\
 & a_5 & \ddots &   & \\
 & & \ddots & 1 &   \\
 & & & a_{2N-1} & 1
\end{pmatrix},
\,
M(\lambda) = 
\begin{pmatrix}
a_2 & 1 & & & \\
&  a_4 & 1 &   & \\
 &  & \ddots & \ddots  & \\
 & & & a_{2N-2} & 1  \\
\lambda & & & & a_{2N}
\end{pmatrix}.
\end{equation}
We denote by $\overline{R}(\lambda), \overline{M}(\lambda)$
the matrices obtained from $R(\lambda), M(\lambda)$ by
replacing $a_i$ by $\bar{a}_i$ for all $i$.
Then
the evolution equations of the discrete periodic Toda lattice 
\eqref{eq:jun30_1} are equivalent to
$\overline{R}(\lambda) \overline{M}(\lambda) = M(\lambda) R(\lambda)$.
Let $L(\lambda) = R(\lambda) M (\lambda), 
\overline{L}(\lambda) =
\overline{R}(\lambda) \overline{M}(\lambda)$.
Then we have the Lax representation for the discrete periodic Toda lattice as
\begin{equation}
\overline{L}(\lambda)  =
M(\lambda) L(\lambda)M (\lambda)^{-1}.
\end{equation}
This implies that the polynomial $\det (x \mathbb{I} + L(\lambda))$ is invariant under the time evolution.
Hence its coefficients are conserved quantities.
Since $G_N(x;a_1,\dots,a_{2N})$ in \eqref{eq:july2_3}
is expressed as $G_N(x;a_1,\dots,a_{2N}) =
\det (x \mathbb{I} + L((-1)^{N-1} y))$
we have the following:
\begin{lemma}\label{le:nov11_6}
The discrete periodic Toda lattice
\eqref{eq:jun30_1} has $N+1$ conserved quantities
\begin{equation*}
h_k = \sum_{\stackrel{1 \leq i_1 \triangleleft \, i_2 \triangleleft \dots \triangleleft \, i_k \leq 2N}{\scriptstyle  (i_1,i_k) \ne (1,2N)}}
a_{i_1} a_{i_2}\dots a_{i_k},
\end{equation*}
for $1 \leq k \leq N$ and $h_{N+1}=\prod_{i=1}^{2N} a_i$.
\end{lemma}
By means of their tropicalization we obtain:
\begin{lemma}\label{le:nov11_6b}
The tropical periodic Toda lattice
\eqref{eq:july2_7} has $N+1$ conserved quantities
\begin{equation}\label{eq:july2_8}
H_k = \min_{\stackrel{1 \leq i_1 \triangleleft \, i_2 \triangleleft \dots \triangleleft \, i_k \leq 2N}{\scriptstyle (i_1,i_k) \ne (1,2N)}}
\left( A_{i_1} + A_{i_2}+ \dots + A_{i_k} \right),
\end{equation}
for $1 \leq k \leq N$ and $H_{N+1}=\sum_{i=1}^{2N} A_i$.
\end{lemma}
\proof
As was explained in \S 4.1 of \cite{IKT12}, the tropicalization can be realized as a concrete limiting procedure.
In the present setting we let $a_i = e^{-A_i/\varepsilon}$ with $\varepsilon > 0$, apply the map
$x \mapsto - \varepsilon \log x$
and take the limit $\varepsilon \to 0$.
This procedure transforms \eqref{eq:jun30_2} to \eqref{eq:july2_7},
which implies the claim of this lemma of the basis of the previous one.
\qed
\begin{remark}
We derived Lemmas \ref{le:nov11_6} and \ref{le:nov11_6b} directly through the Lax representation.
An equivalent result was obtained by using a different method in \cite{IT07}, Proposition 3.9. 
\end{remark}

\subsection{The weak convexity condition relating the conserved quantities}
\label{sec:2_3}
The iso-level set structure of trop p-Toda has been clarified
by means of tropical geometry \cite{IT08, IT09} in case when the {\em strong convexity condition}
$H_k + H_{k+2} > 2 H_{k+1}$ (which they call generic) is satisfied by the conserved quantities.
In this section we prove that in general cases only the {\em weak convexity condition}
$H_k + H_{k+2} \geq 2 H_{k+1}$ holds.
For this purpose we first consider a lemma
in elementary combinatorics.

\begin{figure}[hbtp]
\centering
\scalebox{0.5}[0.5]{
\includegraphics[clip]{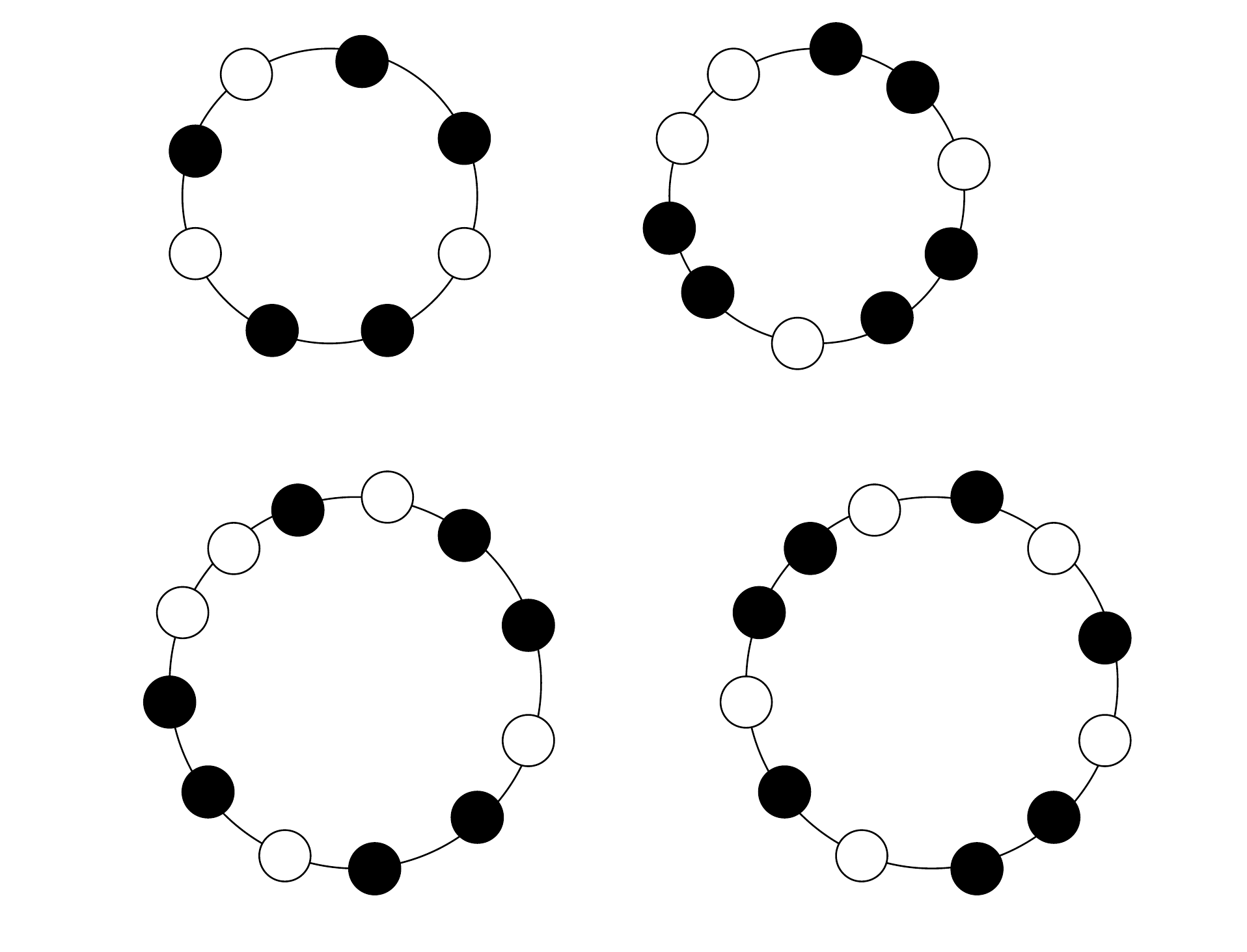}}
\caption{Examples for Lemma \ref{lem:jun1_1}}
\label{figure:1}
\end{figure}

Put two kinds of symbols $\circ$'s and
$\bullet$'s on a circle.
Say two $\circ$'s are {\em adjacent} if
there are no other $\circ$'s between them.
\begin{lemma}\label{lem:jun1_1}
Put $k$ $\circ$'s and $k+2$ $\bullet$'s on a circle so that their positions do not coincide and there are at most two $\bullet$'s between adjacent $\circ$'s.
Then on the circle we always have such a configuration
$  - \bullet - \bullet - (\circ - \bullet)^n - \bullet- $
for some $n \geq 1$.
\end{lemma}
\proof
If there exist more than one $\circ$'s between any adjacent $\bullet$'s on the circle, remove the $\circ$'s until there remains only one.
Suppose the number of $\circ$'s we have removed is $\alpha$.
Since the number of $\circ$ is now $k-\alpha$, the number of the configuration $- \bullet - \bullet -$ on the circle is $(k+2) - (k-\alpha) = \alpha + 2$.
By construction, we have such configurations 
$  - \bullet - \bullet - (\circ - \bullet)^n - \bullet- $ for some $n \geq 1$
between all ($\alpha +2$) adjacent $- \bullet - \bullet -$'s, and
at least two of them will be left unchanged when
we put all ($\alpha$) removed $\circ$'s back into the original positions.
\qed
\begin{example}
See Figure \ref{figure:1}.
\end{example}

Now we present one of our new results in this paper.
\begin{theorem}\label{th:jun10_1}
For $x = (x_1,\dots,x_{2N}) \in (\mathbb{R}_{>0})^{2N}$ let
$H_0(x)=0$ and
\begin{equation}\label{eq:jun10_2}
H_k(x) = 
\min_{\stackrel{1 \leq i_1 \triangleleft \, i_2 \triangleleft \dots \triangleleft \, i_k \leq 2N}{\scriptstyle (i_1,i_k) \ne (1,2N)}}
\left( x_{i_1} + x_{i_2}+ \dots + x_{i_k} \right),
\end{equation} 
for $1 \leq k \leq N$.
Then the relations
$H_k(x) + H_{k+2}(x) \geq 2 H_{k+1}(x)$ 
are satisfied for $0 \leq k \leq N-2$.
\end{theorem}
\proof
Suppose we have $H_k(x) = x_{i_1}+\dots+x_{i_k}$ and 
$H_{k+2}(x) = x_{j_1}+\dots+x_{j_{k+2}}$.
Let $S_1 =\{i_1,\dots,i_k\}$ and $S_2=\{j_1,\dots,j_{k+2}\}$ be the sets of their indices.
To begin with we assume $S_1 \cap S_2 = \emptyset$.
For $a,b\in\{1,\dots,2N\}$ say $b$ is \textbf{next to} $a$ if $|a-b| = 1 \mbox{ or } 2N-1$.
If there exists $b \in S_2$ such that $b$ is not next to $a$
for any $a \in S_1$, then we have
$H_k(x)+x_b \geq H_{k+1}(x)$ and 
$H_{k+2}(x)-x_b \geq H_{k+1}(x)$.
Hence the claim follows.

Suppose otherwise, i.~e.~we assume every $b \in S_2$ is next to some $a \in S_1$.
Draw a circle of circumference $2N$ with a spacial coordinate $1,\dots,2N$ assigned counter-clockwise on it.
Put $\circ$'s on the circle at the positions given by $S_1$,
and put $\bullet$'s at those given by $S_2$. 
Since every $b \in S_2$ is next to some $a \in S_1$,
there exist at most two $\bullet$'s between adjacent $\circ$'s
on the circle.
Hence by Lemma \ref{lem:jun1_1} we have such a configuration 
$  - \bullet - \bullet - (\circ - \bullet)^n - \bullet- $
for some $n \geq 1$ on the circle.
Replace this configuration by
$  - \bullet - \circ - (\bullet - \circ)^n - \bullet-$.
Now let $I_1$ (resp.~$I_2$) be the set of the positions of
$\circ$'s (resp.~$\bullet$'s) on the circle.
Then since $|I_1|=|I_2|=k+1$ we have 
$H_k(x) + H_{k+2}(x) = 
\sum_{i \in I_1} x_i + \sum_{i \in I_2} x_i
\geq 2 H_{k+1}(x)$.

When $S_1 \cap S_2 \ne \emptyset$ but not $S_1 \subset S_2$, we can prove the statement
by replacing $S_i$ by $S_i \setminus (S_1 \cap S_2)$ for
$i=1,2$ and repeating the above arguments.
This is because no elements of $S_1 \cap S_2$
are next to any $a \in S_i \setminus (S_1 \cap S_2)$ for $i=1,2$.
The case $S_1 \subset S_2$ is simpler, because
no elements of $S_2 \setminus S_1$ are next to any $a \in S_1$.
The proof is completed.
\qed

\begin{figure}[h]
\scalebox{1}[1]{
\includegraphics[clip]{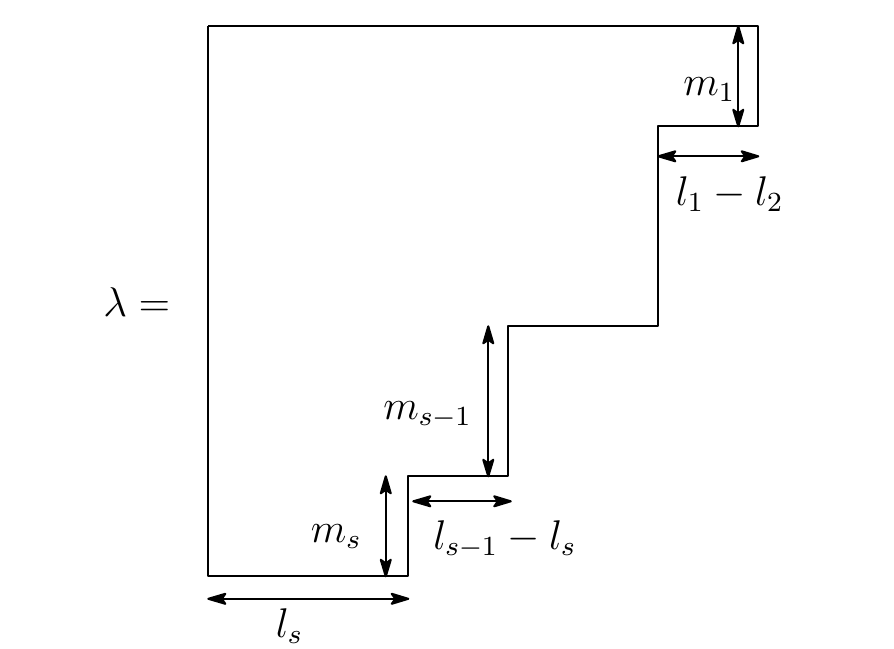}}
\caption{The Young diagram.}
\label{figure:2}
\end{figure}

Recall the conserved quantity of the tropical periodic Toda lattice
\eqref{eq:july2_8}.
Let $\tilde{l}_i = H_{N+1-i}-H_{N-i}$ for $1 \leq i \leq N$.
Then by Theorem \ref{th:jun10_1} we have
$\tilde{l}_1 \geq \dots \geq \tilde{l}_N>0$.
Let $\{ l_i \}_{1 \leq i \leq s}$ be the set of real numbers satisfying $l_1 > \dots > l_s$ such that for any $1 \leq j \leq N$ there exists $1 \leq i \leq s$ such that $\tilde{l}_j = l_i$.
And let $m_i = \# \{ j | 1 \leq j \leq N, \tilde{l}_j = l_i\}$.
Now we can express the conserved quantities by a Young diagram (Figure \ref{figure:2}) in which the lengths of the horizontal edges are
not necessarily integers.

In the context of integrable cellular automaton, this Young diagram represents the content of solitons
in the system.
From this point of view, we have $m_i$ solitons of amplitude $l_i$ for $1 \leq i \leq s$.
We note that the lengths of the black segments in Figure \ref{fig:0} are not always identical to the amplitudes of solitons represented by $\lambda$, since there are intermediate states of collisions of several solitons. 

\subsection{Another algorithm for the Young diagram and the main theorem}\label{sec:2_4}
%
We introduce another algorithm to construct Young diagrams from states of trop p-Toda.
This algorithm is related to the KKR bijection in $\mathfrak{sl}_2$ case
which was applied to the inverse scattering 
transform of pBBS \cite{KTT}.
It is also regarded as a continuous analogue of the `10-elimination' procedure
\cite{YYT}.
We shall prove that the Young diagram obtained here
coincides with the one that was defined in the previous subsection.

%
Say any sequence of nonnegative real numbers
$x_1,\ldots,x_{2n}$ obeys the {\em highest weight condition} if the inequalities
\begin{equation}\label{eq:hwc}
\sum_{i=1}^k (x_{2i-1}-x_{2i}) \geq 0, 
\end{equation}
are satisfied for $1 \leq k \leq n$.
Fix time $t$ and denote the dependent variables of trop p-Toda $A_i^t$ by $A_i$.
By shifting their indices cyclically,
one can make the sequence 
$A_1, \ldots , A_{2N}$ obey the highest weight condition.
This is due to the phase space condition under \eqref{eq:july2_7};
suppose that we have this.

Let $x^{(1)}=(x^{(1)}_1,\ldots,x^{(1)}_{2N^{(1)}})$ with $x^{(1)}_i = A_i$ and $N^{(1)} = N$.
Given an array of positive real numbers $x^{(i)}=(x^{(i)}_1,\ldots,x^{(i)}_{2N^{(i)}})$
satisfying the highest weight condition,
define 
$y^{(i)}=(y^{(i)}_1,\ldots,y^{(i)}_{2N^{(i)}})$
by $y^{(i)}_j = x^{(i)}_j - \mu^{(i)}$ where $\mu^{(i)}= \min_{1 \leq j \leq 2N^{(i)}}\{x^{(i)}_j \} $.
In the array of non-negative real numbers
$y^{(i)}_1,\ldots,y^{(i)}_{2N^{(i)}}$, suppose there are $k^{(i)}$ sequences of zeros.
Here we regard a lone zero also as a sequence.
We denote by $n_j^{(i)} \, (1 \leq j \leq k^{(i)})$ the length of the $j$th sequence of zeros.
Let $N^{(i+1)}=N^{(i)} - \sum_{j=1}^{k^{(i)}} \lceil n_j^{(i)}/2 \rceil$ where $\lceil c \rceil$ is the smallest integer satisfying $\lceil c \rceil \geq c$.
If $N^{(i+1)} = 0$ then we stop. 
Otherwise we define an array
$x^{(i+1)}=(x^{(i+1)}_1,\ldots,x^{(i+1)}_{2N^{(i+1)}})$ 
of positive real numbers satisfying the highest weight condition by the following procedure.

\begin{figure}[h]
\scalebox{1}[1]{
\includegraphics[clip]{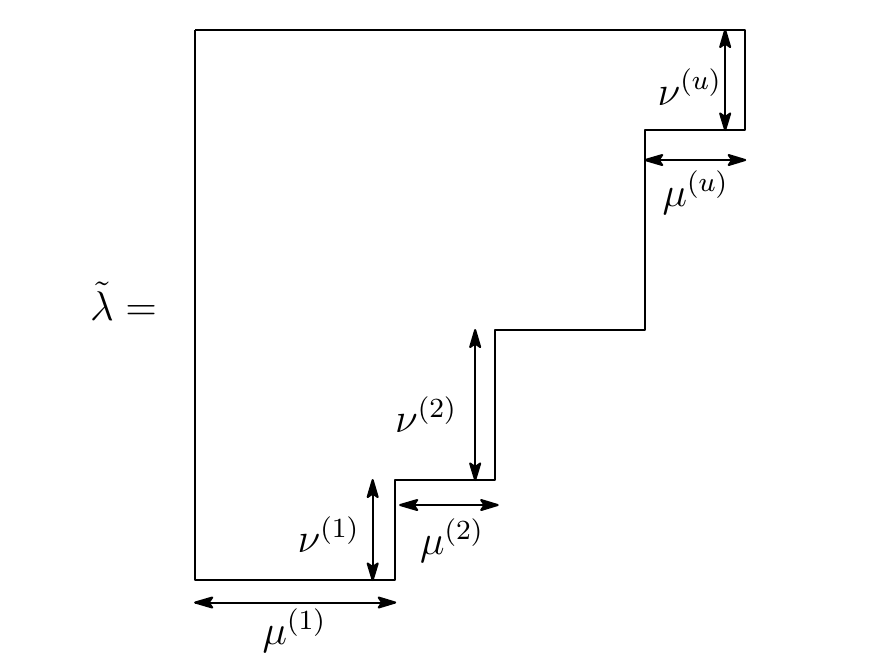}}
\caption{The Young diagram.}
\label{figure:3}
\end{figure}

\begin{enumerate}
\item \label{item:dec18_1}
If the first
sequence of $n_1^{(i)}$ zeros 
in the array $y^{(i)}_1,\ldots,y^{(i)}_{2N^{(i)}}$
is at the left end, then
remove these zeros.
We see that $n_1^{(i)}$ must be even because
the array satisfies the highest weight condition
\item
For any $j$ such that the $j$-th sequence of $n_j^{(i)}$ zeros 
is between positive neighbors $a,b$
as $\ldots,a,0,\ldots,0,b,\ldots$,
replace it by $\ldots, a, b, \ldots$ if $n_j^{(i)}$ is even, or by $\ldots, a+b, \ldots$ if $n_j^{(i)}$ is odd.
More precisely, we simply remove $n_j^{(i)}$ zeros
in the former, or in the latter we further replace the neighbors $a,b$ by a single number $a+b$. 
\item \label{item:dec18_3}
Suppose the last
sequence of $n_{k^{(i)}}^{(i)}$ zeros 
is at the right end 
after a positive neighbor $a$ as
$\ldots,a,0,\ldots,0$.
Remove these zeros.
If $n_{k^{(i)}}^{(i)}$ is odd then also remove 
$a$, and add it to the first element of the array.
\end{enumerate}
Let $x^{(i+1)}_j$ be the $j$th element of the resulting array of $2N^{(i+1)}$ positive integers.
By the following lemma we see that the sequence
$x^{(i+1)}_1,\ldots,x^{(i+1)}_{2N^{(i+1)}}$ satisfies the highest weight condition.
\begin{lemma}\label{lem:jul14_1x}
The highest weight condition is preserved
under the following procedures.
\begin{enumerate}
\item Insert or remove two consecutive zeros.
\item Split any positive term into two positive numbers and insert a zero between them, or remove a zero 
and join its neighbors into one term.
\end{enumerate} 
\end{lemma}
Suppose $N^{(i)}>0$ for $1 \leq i \leq u$ and $N^{(u+1)}=0$.
Let $\nu^{(i)} = N^{(i)} - N^{(i+1)}$.
Obviously the set of numbers 
$\{ (\mu^{(i)},\nu^{(i)}) \}_{1 \leq i \leq u}$
determines a Young diagram $\tilde{\lambda}$
in which the lengths of the horizontal edges are positive real numbers (Figure \ref{figure:3}).
\begin{example}\label{ex:nov17_1}
For $N=6$, let $A_1=2\sqrt{2}, A_2 = A_{12}=2, A_3 = A_8 = A_{11} = \sqrt{2}, A_4=A_5=A_6=A_7=A_9=A_{10}=1$.
Then we have:
\begin{align*}
x^{(1)} &= (2\sqrt{2},2,\sqrt{2},1,1,1,1,\sqrt{2},1,1,\sqrt{2},2),\\
y^{(1)} &= (2\sqrt{2}-1,1,\sqrt{2}-1,0,0,0,0,\sqrt{2}-1,0,0,\sqrt{2}-1,1),\\
x^{(2)} &= (2\sqrt{2}-1,1,\sqrt{2}-1,\sqrt{2}-1,\sqrt{2}-1,1),\\
y^{(2)} &= (\sqrt{2},2-\sqrt{2},0,0,0,2-\sqrt{2}),\\
x^{(3)} &= (\sqrt{2},4-2\sqrt{2}),\\
y^{(3)} &= (3\sqrt{2}-4,0),
\end{align*}
and $\mu^{(1)}=1,\mu^{(2)}=\sqrt{2}-1,\mu^{(3)}=4 -2\sqrt{2},\nu^{(1)}=3,\nu^{(2)}=2,\nu^{(3)}=1$.
\end{example}
\begin{example}\label{ex:dec22_1}
For $N=4$, let $A_1=4, A_2 = A_3 = 3, A_4=A_5=2, A_6=A_7=A_8=1$.
Then we have:
\begin{align*}
x^{(1)} &= (4,3,3,2,2,1,1,1),\\
y^{(1)} &= (3,2,2,1,1,0,0,0),\\
x^{(2)} &= (4,2,2,1),\\
y^{(2)} &= (3,1,1,0),\\
x^{(3)} &= (4,1),\\
y^{(3)} &= (3,0),
\end{align*}
and $\mu^{(1)}=\mu^{(2)}=\mu^{(3)}=1,
\nu^{(1)}=2,\nu^{(2)}=\nu^{(3)}=1$.
\end{example}

Now we present the main result of this paper.
\begin{theorem}\label{th:main}
For any integer $1 \leq k \leq N$,
the area of the part of the Young diagram $\tilde{\lambda}$ between its bottom line and
the horizontal line above it at height $k$ 
is given by $H_k$, the $k$th conserved quantity of the
trop p-Toda defined as in \eqref{eq:july2_8}.
\end{theorem}
In other words, the Young diagram $\tilde{\lambda}$ in Figure \ref{figure:3} coincides with the Young diagram $\lambda$ in 
Figure \ref{figure:2}.
We shall give a proof of this theorem in the next section.

\section{Proof of the main theorem}\label{sec:3}
\subsection{Drawing a diagram of trees}\label{sec:3_1}
\begin{figure}[hbtp]
\centering
\scalebox{0.5}[0.5]{
\includegraphics[clip]{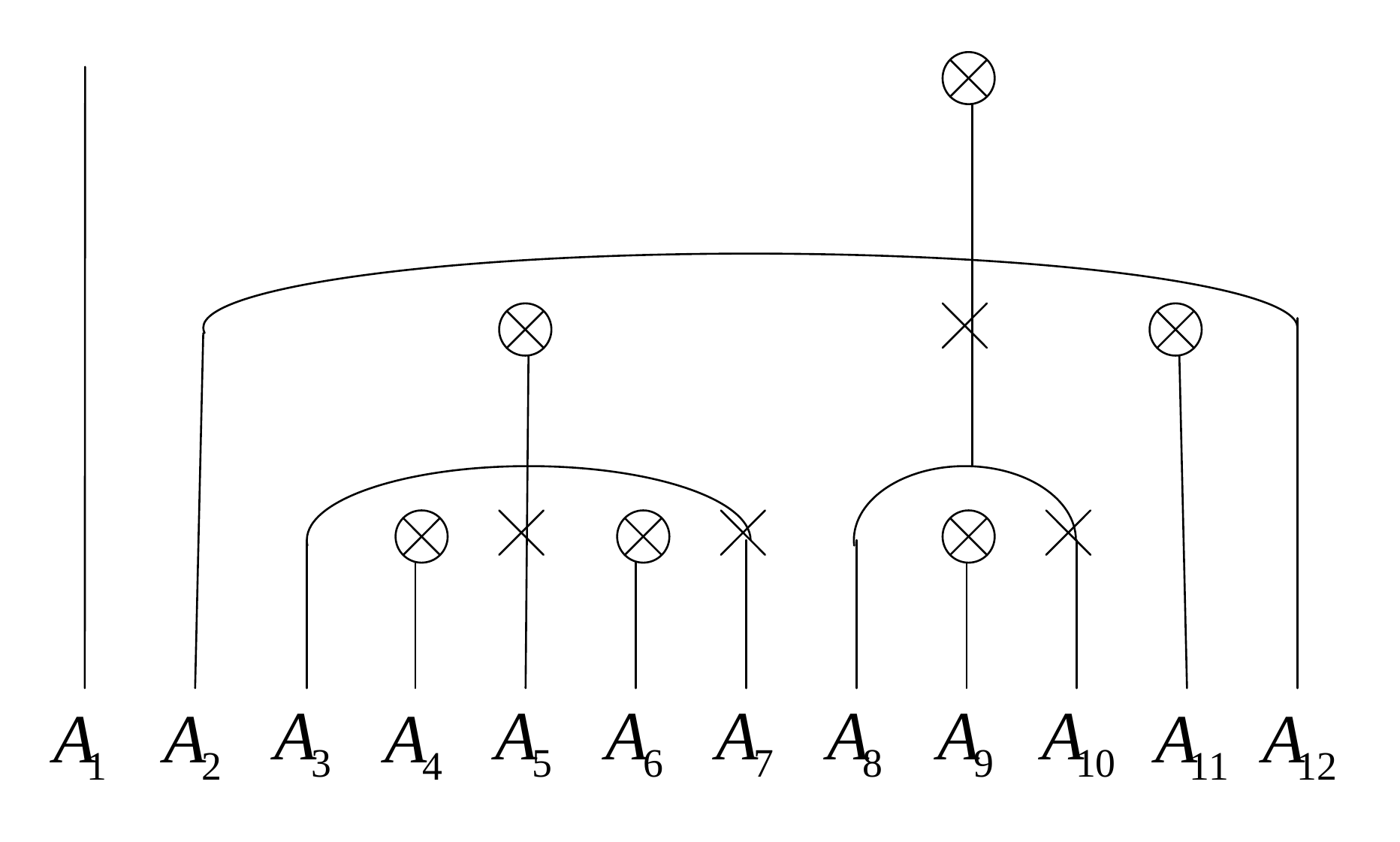}}
\caption{A tree diagram for the $A$-variables}
\label{figure:4}
\end{figure}

By generalizing the method of Appendix A.~1 of \cite{IT07},
we introduce a way to draw a graph $\Phi_A$ associated with a state of the trop p-Toda $A = (A_1,\ldots,A_{2N})$.
It has several connected components called trees.
See Figure \ref{figure:4} for an example, that is for the $A$-variables in Example \ref{ex:nov17_1}.
At the end we rewrite the assertion of Theorem \ref{th:main} in terms of the trees.

Recall the algorithm in \S \ref{sec:2_4}
where $N = N^{(1)}$.
Place the symbols $A_1,\ldots,A_{2N}$
or
$x^{(1)}_1,\ldots,x^{(1)}_{2N^{(1)}}$
at a horizontal level, called level $0$.
Draw a vertical line from each symbol upwardly
to a certain horizontal level, called level $1$.
We associate
the non-negative real numbers
$y^{(1)}_1,\ldots,y^{(1)}_{2N^{(1)}}$ to the lines.
If $y^{(1)}_j=0$, we put a symbol $\times$ at the top of the line.
Then, change the symbol $\times$ into another symbol $\otimes$ if it is an isolated one, or is 
at an `odd-th' position of a sequence of consecutive $\times$'s.
In what follows we will pay attention to 
$\otimes$'s only and ignore the other $\times$'s.
Each $\otimes$ is called the top of a tree of level $1$.
For each isolated $\otimes$ or
sequence of $\otimes$'s placed at
every other position,
we let the lines adjacent to $\otimes$'s
join together to straddle the $\otimes$'s.
Here we respect the periodic boundary condition, so the leftmost and rightmost ends are regarded as adjacent.
We call each joining point a branching point
of level $1$.
After this procedure, the number of lines is reduced to $2 N^{(2)}$.

Now we describe a general procedure
to draw the diagram
from level $i-1$ to level $i$.
We associate
the positive real numbers
$x^{(i)}_1,\ldots,x^{(i)}_{2N^{(i)}}$ to the tops of the $2N^{(i)}$ lines at level $i-1$.
Extend the lines upwardly
to level $i$.
We associate
the non-negative real numbers
$y^{(i)}_1,\ldots,y^{(i)}_{2N^{(i)}}$ to the lines.
If $y^{(i)}_j=0$, we put a symbol $\times$ at the top of the line.
By repeating the same procedure in the previous paragraph, we obtain the tops of the trees, as 
well as the branching points, of level $i$.

At the end we obtain a graph $\Phi_A$ 
for the state $A = (A_1,\ldots,A_{2N})$.
Denote by $\Trees (\Phi_A)$ the set of all trees in the graph $\Phi_A$.
Define the level of a tree by the level of its top point marked by $\otimes$.
By construction, the number of level $i$ trees
is $\nu^{(i)}$.
Hence there are $\sum_{l=1}^u \nu^{(l)} = N$ trees in total.
Let $t \in \Trees (\Phi_A)$ be a tree of level $i$.
We define its height by
${\rm Ht} (t) = \sum_{l=1}^i \mu^{(l)}$.
\begin{definition}\label{def:july4_1}
We label all the trees in $\Trees (\Phi_A)$
as $t_1,\dots,t_N$ so that their heights are in weakly increasing order, i.~e.~$i < j \Rightarrow {\rm Ht}(t_i) \leq {\rm Ht} (t_j)$.
Let $\Tree^{(k)}= \{ t_1,\ldots,t_k\}$.
\end{definition}
Now we see that the assertion of Theorem \ref{th:main}
is equivalent to the relation
\begin{equation}\label{eq:nov18_1}
\sum_{i=1}^k {\rm Ht}(t_i) = \min_{\stackrel{1 \leq i_1 \triangleleft \, i_2 \triangleleft \dots \triangleleft \, i_k \leq 2N}{\scriptstyle (i_1,i_k) \ne (1,2N)}}
\left( A_{i_1} + A_{i_2}+ \dots + A_{i_k} \right),
\end{equation}
for $1 \leq k \leq N$.

\subsection{Elementary lemmas}\label{sec:3_2}
We present several elementary lemmas
on the graph $\Phi_A$
that are necessary for proving Theorem \ref{th:main}.
%
Given $t \in \Trees (\Phi_A)$, we define
a set of indices of the $A$-variables as
\begin{equation}\label{eq:nov29_1}
\Root (t) = \{
i \in \{1,\dots,2N\} \vert 
\mbox{$A_i$ is at a bottom point of $t$} \}.
\end{equation}
For $T \subset \Trees (\Phi_A)$ we write
$\Root (T) = \bigsqcup_{t \in T} \Root (t)$.
If $P$ is a branching point of $t$ at level $i$,
we define its height by
${\rm Ht} (P) = \sum_{l=1}^i \mu^{(l)}$.
(Note that we regard a tree and a branching point of the same level as having common height, though they are not so depicted in figures for technical reasons.)
Say $P$ has multiplicity $m_P$ if there are 
$m_P + 2$ lines outgoing from $P$.
If the tree $t$ has branching points $P_1,\dots, P_q$ with multiplicities $m_{P_1},\dots, m_{P_q}$,
define the weight of $t$ by
\begin{equation}
\wt (t) = {\rm Ht}(t) + \sum_{j=1}^q {\rm Ht}(P_j) m_{P_j}.
\end{equation}
Then we have:
\begin{lemma}[\cite{IT07}, Lemma A.1]\label{lem:nov22_1}
For any $t \in \Trees (\Phi_A)$ it holds that
\begin{equation}
\wt (t) = \sum_{i \in \Root (t)} A_i.
\end{equation}
\end{lemma}
The set $\Trees (\Phi_A)$ becomes a partially ordered set
on introducing the following partial order. 
Denote by $t \succ s$
when $s$ is straddled by $t$.
We denote by $t \succeq s$
a case where either $t \succ s$ or $t=s$ is satisfied.
For $t \in \Trees (\Phi_A)$ we define
\begin{align}\label{eq:nov29_2}
\Sub (t)&= \{ s \in \Trees (\Phi_A) \vert 
t \succ s \},\\
\oSub (t)&= \{ s \in \Trees (\Phi_A) \vert 
t \succeq s \}= \Sub (t) \sqcup \{ t \}.\label{eq:dec22_3}
\end{align}
Each element of $\oSub (t)$ takes either
an `odd-th' position or an `even-th' position
in its nesting structure,
regarding
$t$ itself  as taking the first position and the order of the nesting as increasing inwardly.
We denote by $\oSubo (t) \subset \oSub (t)$ the set of all trees
at odd-th positions, and let
$\oSube (t) = \oSub (t) \setminus \oSubo (t)$.
Accordingly, we define
\begin{align}
\oRooto (t) &=  
\Root (\oSubo (t)),\\
\oRoote (t) &= 
\Root (\oSube (t)),
\end{align}
and $\oRoot (t) = \oRooto (t) \sqcup \oRoote (t)$.
Then we have:
\begin{lemma}[\cite{IT07}, Lemma A.2]\label{lem:nov21_1}
For any $t \in \Trees (\Phi_A)$ it holds that
\begin{equation}
\sum_{s \in \oSub (t)} {\rm Ht} (s) =
\sum_{i \in \oRooto (t)} A_i.
\end{equation}
\end{lemma}
We also use the following lemmas in the next subsection.
\begin{lemma}\label{lem:nov21_2}
\begin{equation}
\# \oSub (t) =
\# \oRooto (t).
\end{equation}
\end{lemma}
\proof
Denote all the branching points in $\oSub (t)$
by $P_1,\dots, P_r$ and their multiplicities by $m_{P_1},\dots, m_{P_r}$.
By looking the graph downwardly, we see that the number of trees increases by $m$ at a branching point with multiplicity $m$.
Hence $\# \oSub (t) = 1 + \sum_{i=1}^r m_{P_i}$.
In the same way, 
we see that the number of vertical lines increases by $2m$ at the branching point.
Hence $\# \oRoot (t) = 1 + \sum_{i=1}^r 2m_{P_i}$, which implies 
$\# \oRooto (t) = 1 + \sum_{i=1}^r m_{P_i}$.
\qed

\vspace{5mm}
Recall the definition of the set $\Tree^{(k)}$ in Definition \ref{def:july4_1}.
One can prove the following lemma by induction on $k$.
\begin{lemma}\label{lem:nov21_3}
For any $t \in \Tree^{(k)}$
the relation $\oSub (t) \subset \Tree^{(k)}$
holds.
\end{lemma}

Say $\xi$ is a maximal element of a partially ordered set $X$ if $\xi \succeq \xi'$ is satisfied for any $\xi'  \in X$ 
that is {comparable} to $\xi$ with respect to the partial order.
\begin{definition}
Let $\MTree^{(k)} \subset \Tree^{(k)}$
be the set of all maximal elements of $\Tree^{(k)}$.
\end{definition}
Then we have:
\begin{lemma}\label{lem:nov21_4}
\begin{equation}
\bigsqcup_{t \in \MTree^{(k)}} \oSub (t) =
\Tree^{(k)}.
\end{equation}
\end{lemma}
\proof
We show the inclusion $\subset$ since
the opposite inclusion is almost trivial.
Suppose $u$ is an element of LHS.
Then there exists $t \in \MTree^{(k)} \subset \Tree^{(k)}$ such that $u \in \oSub (t)$.
This implies $u \in \Tree^{(k)}$
by Lemma \ref{lem:nov21_3}.
\qed
\begin{definition}
Let $\MSub (t) \subset \Sub (t)$
be the set of all maximal elements of $\Sub (t)$.
\end{definition}
Then it is easy to see that the following relations are satisfied:
\begin{align}
\Sub (t) &= \bigsqcup_{s \in \MSub (t)} \oSub (s),\label{eq:nov22_1}\\
\oRoote (t) &= \bigsqcup_{s \in \MSub (t)} \oRooto (s). \label{eq:nov22_2}
\end{align}

From \eqref{eq:dec22_3}, \eqref{eq:nov22_1}, \eqref{eq:nov22_2} and Lemma \ref{lem:nov21_1} we have:
\begin{lemma}\label{lem:nov22_2}
\begin{equation}
\sum_{i \in \oRooto (t)} A_i =
{\rm Ht} (t) + \sum_{i \in \oRoote (t)} A_i.
\end{equation}
\end{lemma}

\begin{figure}[hbtp]
\centering
\scalebox{0.5}[0.5]{
\includegraphics[clip]{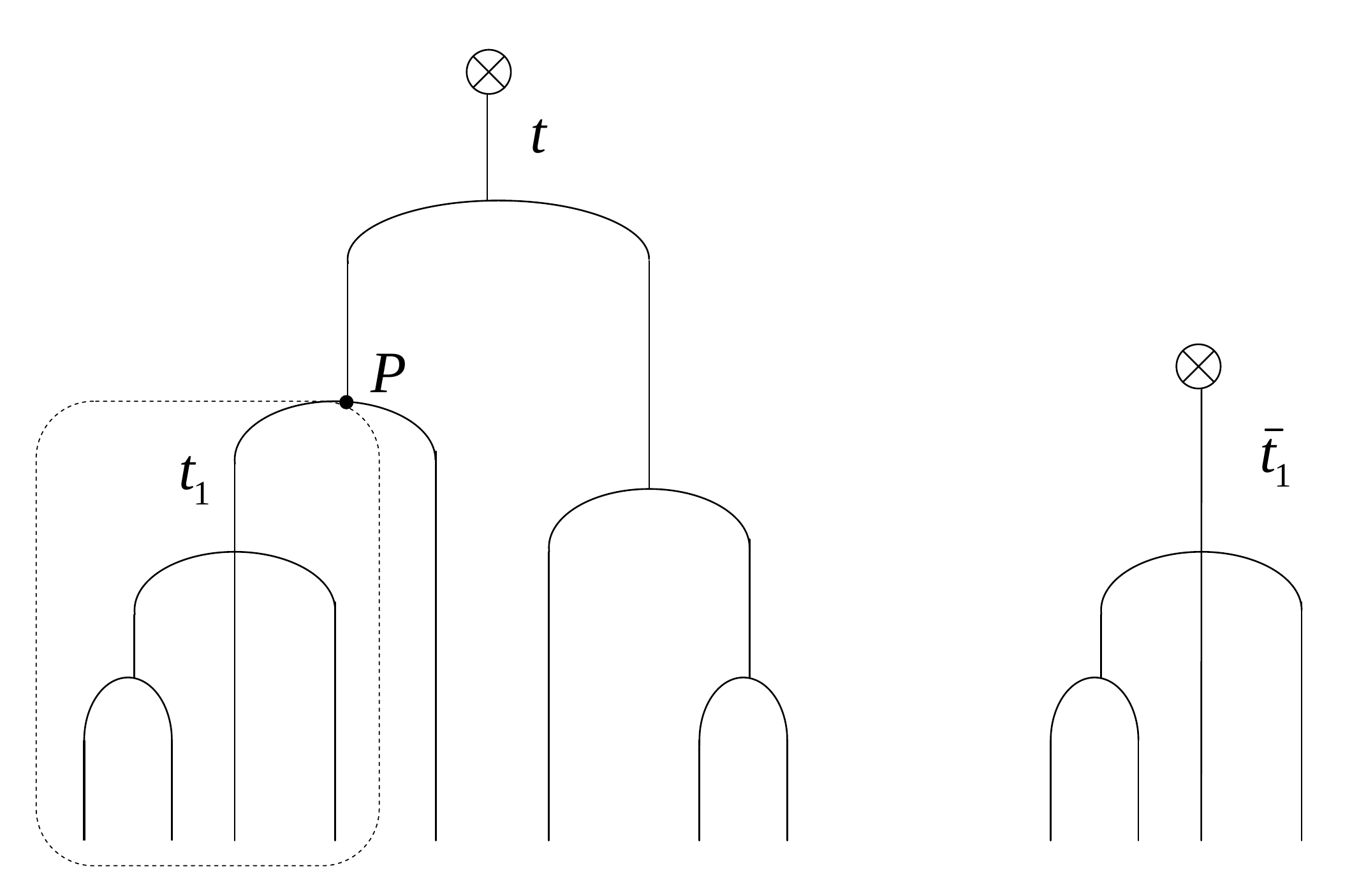}}
\caption{(left) The subtree $t_1$ defined as a branch of $t$ under a branching point $P$: (right) A tree $\overline{t}_1$ defined as the completion of $t_1$}
\label{figure:5}
\end{figure}

Let $t \in \Trees (\Phi_A)$ be a tree and $P \in t$ is one of its branching points.
Consider a subtree $t_1 \subset t$ that extends downwardly from $P$.
See Figure \ref{figure:5}.
Define $\Root (t_1), \Sub (t_1)$, etc.~by
extending the definitions \eqref{eq:nov29_1},
\eqref{eq:nov29_2}, etc.~in an obvious way.
Note that $t_1$ itself is not an element of $\Trees (\Phi_A)$.
Then we have:
\begin{lemma}\label{lem:nov22_3}
\begin{equation}
\sum_{i \in \oRooto (t_1)} A_i \geqq
{\rm Ht} (P) + \sum_{i \in \oRoote (t_1)} A_i.
\end{equation}
\end{lemma}
\proof
If the subtree $t_1$ has branching points $Q_1,\dots, Q_s$ with multiplicities $m_{Q_1},\dots, m_{Q_s}$,
define its weight by
\begin{equation}
\wt (t_1) = {\rm Ht}(P) + \sum_{j=1}^s {\rm Ht}(Q_j) m_{Q_j}.
\end{equation}
Then by the algorithm of drawing the graph $\Phi_A$ we can deduce
\begin{math}
\wt (t_1) \leq \sum_{i \in \Root (t_1)} A_i.
\end{math}
We define $\overline{t}_1$
called the completion of $t_1$
as a tree obtained
by extending the top of $t_1$ by the length
$\sum_{i \in \Root (t_1)} A_i - \wt (t_1)$.
In other words $\overline{t}_1$ is a tree that shares all the branching/bottom points with $t_1$ but satisfies Lemmas \ref{lem:nov22_1} and \ref{lem:nov21_1}.
Then the claim follows by applying Lemma \ref{lem:nov22_2} on the tree $\overline{t}_1$ and using $\wt (\overline{t}_1) \geq \wt (P)$.
\qed

\subsection{Proof of Theorem \ref{th:main}}\label{sec:3_3}
Now we give a proof of the main theorem, leaving
proofs of two more lemmas to appear
afterwards in the following subsections.
 
Given $N$ we define the sets of 
nearest neighbor excluding
indices as
\begin{align}
{\mathcal B}(k,N) &=
\{ \{ i_1,\ldots,i_k \} \subset \Z^k \vert
1 \leq i_1 \triangleleft i_2 \triangleleft \dots \triangleleft i_k \leq 2N, (i_1,i_k) \ne (1,2N)\},\\
{\mathcal B}(N) &= \bigsqcup_{1 \leqq k \leqq N}{\mathcal B}(k,N).
\end{align}
Then the RHS of \eqref{eq:nov18_1}
can be written as $\min_{B \in {\mathcal B}(k,N)} \{ \sum_{i \in B} A_i \}$.
The {\em Forest Realization Lemma}
in \S \ref{subsec:FRL} claims that
for any $1 \leq k \leq N$
there exists $T \subset \Trees (\Phi_A)$ such that
the relation
\begin{equation}\label{eq:nov21_2}
\min_{B \in {\mathcal B}(k,N)} \left\{ \sum_{i \in B} A_i \right\} = \sum_{i \in  \Root (T)} A_i,
\end{equation}
is satisfied.
Then, from the {\em Close Packing Lemma}
in \S \ref{subsec:CPL}
we can deduce that such $T$ must be written 
by a disjoint union as
\begin{equation}
T = \bigsqcup_{t \in U} \oSubo (t),\quad
U \subset \Trees (\Phi_A).
\end{equation} 
This implies that
\begin{equation}
\sum_{i \in  \Root (T)} A_i =
\sum_{t \in U} \sum_{i \in \oRooto (t)} A_i =
\sum_{t \in U} \sum_{s \in \oSub (t)} {\rm Ht} (s)
\geq \sum_{i=1}^k {\rm Ht} (t_i),
\end{equation}
where we have used Lemma \ref{lem:nov21_1}
and the relation $\sum_{t \in U} \# (\oSub (t))=k$
which was verified by Lemma \ref{lem:nov21_2}.
Hence it suffices to show that there exists
$U \subset \Trees (\Phi_A)$ such that the relation
\begin{equation}
\bigsqcup_{t \in U} \oSub (t) = \{ t_1,\ldots,t_k\},
\end{equation}
is satisfied.
By Lemma \ref{lem:nov21_4} one finds that this relation is satisfied when $U = \MTree^{(k)}$.
This completes the proof of Theorem \ref{th:main}.

\subsection{Close Packing Lemma}\label{subsec:CPL}\label{sec:3_4}
Recall that ${\mathcal B}(N) = \bigsqcup_{1 \leqq k \leqq N}{\mathcal B}(k,N)$
is the set of nearest neighbor excluding indices.
Given $t \in \Trees (\Phi_A)$,
take any
$V \subset \oSub (t)$ satisfying
$t \in V$ and $\Root (V) \in {\mathcal B}(N)$.
We say that $V$ is {\em closely packed 
with respect to $t$}
when $V = \oSubo (t)$. 

\begin{lemma}\label{lem:CPL}
If $V$ is not closely packed with respect to $t$, there exists
$S \in {\mathcal B}(N)$
such that $S \subset  \oRoot (t), |S| = |\Root (V)|$ and
$\sum_{i \in S} A_i < \sum_{i \in \Root (V)} A_i$.
\end{lemma}
\proof
Let $n$ be the order of the nesting of the trees in $\oSub (t)$.
We prove the lemma by induction on $n$.
If $n=1,2$ one necessarily has $V = \oSubo (t)$ so there is nothing to be proved.
Suppose $n \geq 3$.
Let $V^C = \oSub (t) \setminus V$ and
$a = \min  \Root (\oSubo (t) \cap V^C) $.
If such $a$ does not exist, then $V$ is closely packed.
Suppose otherwise.
We denote by $s \in \oSubo (t) \cap V^C$ the tree
satisfying $a \in \Root (s)$.

\begin{figure}[hbtp]
\centering
\scalebox{0.5}[0.5]{
\includegraphics[clip]{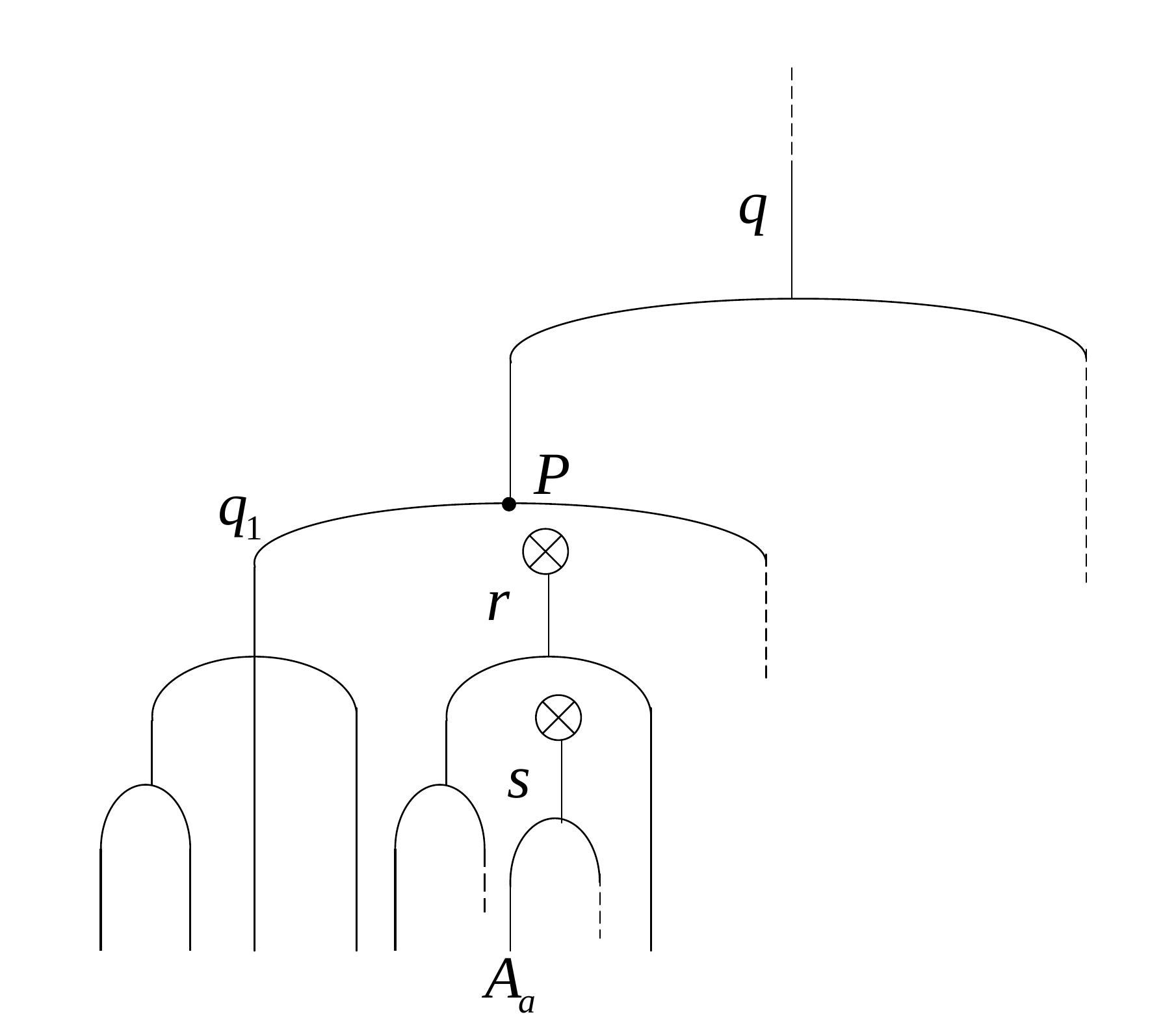}}
\caption{Nested trees $q,r,s$ for the proof of Lemma \ref{lem:CPL}, case  \ref{case2a}.}
\label{figure:6}
\end{figure}

\begin{enumerate}
\item \label{case1}
Suppose there exists $u \in \MSub (s)$ such that both
$u \in V$ and
$\oSubo (u) \cap V \ne \oSubo (u) $ are satisfied.
Then $V_u = V \cap \oSub (u)$ is not closely packed with respect to $u$.
From the induction hypothesis,
there exists
$S_u \in {\mathcal B}(N)$
such that $S_u \subset  \oRoot (u), |S_u| = | \Root (V_u)|$ and
$\sum_{i \in S_u} A_i < \sum_{i \in \Root (V_u)} A_i$.
Now the assertion of the lemma follows on taking
$S = (\Root (V) \setminus \Root (V_u)) \sqcup S_u$.

\item \label{case2}
Suppose otherwise, i.~e.~for any $u \in \MSub (s)$
either $u \notin V$ or $\oSubo (u) \subset V$ is
satisfied.

\begin{enumerate}
\item \label{case2a}
Suppose $\oSubo (u) \subset V$ is satisfied for any
$u \in \MSub (s)$.
This implies that
$\oSube (s) \subset V$
or equivalently $\oRoote (s) \subset \Root (V)$.
Let $r \in  \oSube (t)$ be the tree that directly straddles $s$,
and $q \in  \oSubo (t)$ be the one that directly straddles $r$.
See Figure \ref{figure:6}.
We denote by $P$ the branch point of $q$ at which it straddles $r$.
Note that ${\rm Ht} (P) = {\rm Ht} (r)$.
Let $q_1$ be the subtree of $q$ that extends 
downwardly from $P$
and is adjacent to $r$ on its left side.
By the definition of $a$ we have
$\oRooto (q_1) \subset \Root (V)$.
Define $S \in {\mathcal B}(N)$ as
\begin{equation}
S = \{ \Root (V) \setminus (\oRooto (q_1) \sqcup \oRoote (s)) \} \sqcup (\oRoote (q_1) \sqcup \oRooto (s)).
\end{equation}
Then by Lemmas \ref{lem:nov22_2} and \ref{lem:nov22_3}
we have
$\sum_{i \in \Root (V)} A_i - \sum_{i \in S} A_i  \geq {\rm Ht} (P) - {\rm Ht} (s) > 0$.
\item \label{case2b}
Suppose otherwise, i.~e.~ there exists $s' \in \MSub (s)$ such that $s' \notin V$.
Replace $s$ by $s'$ and repeat the above arguments.
Since the order of the nesting is finite, this case \ref{case2b}
cannot repeat endlessly, and we will eventually arrive at
case \eqref{case1} or \ref{case2a}. 
The proof is completed.
\end{enumerate}
\end{enumerate}

\qed
\subsection{Forest Realization Lemma}\label{subsec:FRL}\label{sec:3_5}
Given a state of the trop p-Toda $A = (A_1,\ldots,A_{2N})$,
there are generally more than one
$B^*\in {\mathcal B}(k,N)$ which satisfy the condition 
\begin{math}
\min_{B \in {\mathcal B}(k,N)} \left\{ \sum_{i \in B} A_i \right\} = \sum_{i \in B^*} A_i.
\end{math}
We want to show that it is always possible to 
find such $B^*$ that can be realized as the set of all the bottom points of a {\em forest},  
a set of trees in $\Phi_A$.

\begin{example}\label{ex:jan11_1}
Consider Example \ref{ex:nov17_1}
with $k=4$.
One can take $B^* = \{ 4,6,8,10 \}, \{5,7,9,11\}, \{3,5,7,9\}$
or $\{4,6,9,11\}$.
In Figure \ref{figure:4} one finds that the former two are not realized by forests, but the latter two are.
\end{example}

\begin{lemma}\label{lem:FRL}
For any $1 \leq k \leq N$
there exists $T \subset \Trees (\Phi_A)$ 
such that both 
$\Root (T) \in {\mathcal B}(k,N)$ and 
\begin{equation}\label{eq:nov25_1}
\min_{B \in {\mathcal B}(k,N)} \left\{ \sum_{i \in B} A_i \right\} = \sum_{i \in \Root (T)} A_i,
\end{equation}
are satisfied.
\end{lemma}

\proof
Recall that $\Phi_A$ is a graph associated with $A = (A_1,\dots,A_{2N})$.
It is composed of several connected components called trees.
We denote by $N (\Phi_A)$ the set of all nodes of $\Phi_A$.
It is the set of top points, bottom points, and branching points of the trees in $\Phi_A$.
In the same way, we denote by $L (\Phi_A)$ the set of all links of $\Phi_A$.
Note that, a top point has only a downward link, a bottom point has only an upward link, and a branching point has an upward and several downward links outgoing from it.

Choose any
$B^* \in {\mathcal B}(k,N)$ that satisfies 
\begin{equation}\label{eq:nov27_1}
\min_{B \in {\mathcal B}(k,N)} \left\{ \sum_{i \in B} A_i \right\} = \sum_{i \in B^*} A_i.
\end{equation}
We draw a subgraph of $\Phi_A$ associated with $B^*$, that is denoted by $\Phi_A (B^*)$ and is defined as follows.
First we adopt the bottom points $\{ A_i \}_{i \in B^*}$ as elements of $N(\Phi_A (B^*))$,
and adopt the links connected to them as
those of $L(\Phi_A (B^*))$.
We also adopt the nodes at the other end of these links as elements of $N(\Phi_A (B^*))$.
If such an adopted node is a branching point of $\Phi_A$, then there are two cases to be distinguished.
\begin{enumerate}
\item Filled branching point: all its downward links are adopted ones.
\item Unfilled branching point: some of its downward links are unadopted ones.
\end{enumerate}
If there is a filled branching point, we also adopt its upward link and the node at the other end as elements of $L(\Phi_A (B^*))$ and $N(\Phi_A (B^*))$.
Repeat this procedure as much as possible, and let $\Phi_A (B^*)$ be the graph 
obtained finally. 
If all the branching points in $\Phi_A (B^*)$ are filled ones, then there exists $T \subset \Trees (\Phi_A)$ 
such that $B^* = \Root (T)$.
Hence we are done.

Suppose otherwise.
It is enough to show that 
there exists a procedure for finding a
$C^* \in {\mathcal B}(k,N)$ 
such that
the number of unfilled branching points 
in $\Phi_A (C^*)$ is smaller than the number of those
in $\Phi_A (B^*)$ by one, under the condition
\begin{math}
\sum_{i \in C^*}A_i = \sum_{i \in B^*}A_i
\end{math}.
The claim of the lemma follows
from using this procedure repeatedly.
%
Now we start describing such a procedure.
We denote by $P$ an arbitrary chosen
unfilled branching point in $\Phi_A (B^*)$ with lowest height, and by $q \in \Tree (\Phi_A)$ the 
tree wherein $P$ lies.
By definition, 
there are both adopted and unadopted subtrees of $q$ under $P$.
It is enough to show that 
there exists a way to reduce the number of
the adopted subtrees
without changing the other conditions.

\begin{figure}[hbtp]
\centering
\scalebox{0.5}[0.5]{
\includegraphics[clip]{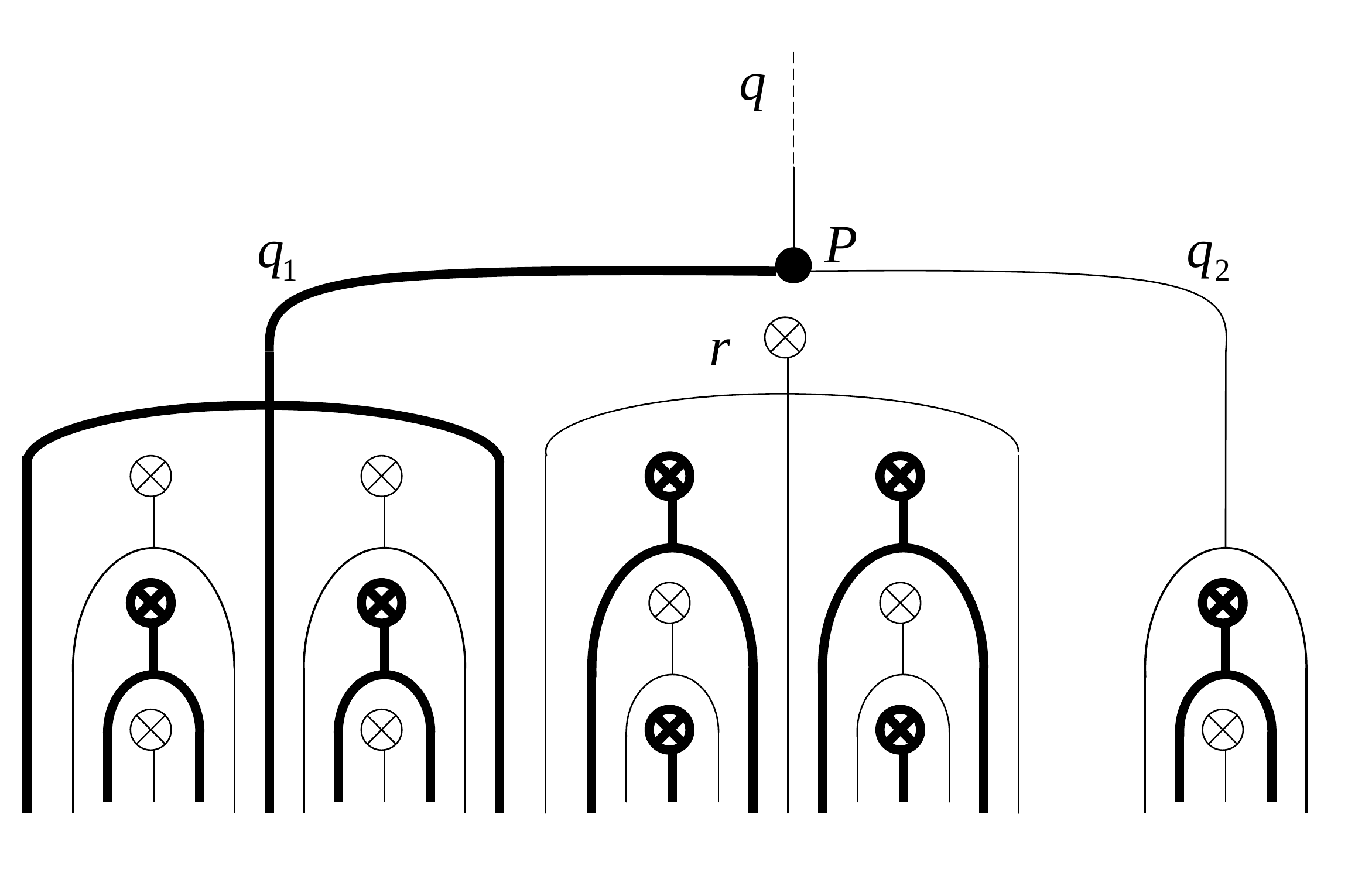}}
\caption{An unfilled branching point $P$ lies in the tree $q$. Thick lines are adopted links, and thin lines are unadopted ones.}
\label{figure:7}
\end{figure}

Among those subtrees, choose an adjacent pair of adopted/unadopted subtrees $q_1, q_2$. 
See Figure \ref{figure:7} for an example.
Then, there is an unadopted 
tree $r \in \MSub (q)$ under $P$ between $q_1$ and $q_2$.
By Lemma \ref{lem:CPL} 
and since there is no unfilled branching point
under $P$
one sees that
$\oSub (q_1) \cap \Phi_A (B^*)$
must be closely packed with respect to $q_1$,
because otherwise 
\eqref{eq:nov27_1} is not satisfied.

Let
$B^*_1 = (\oRoot ({q}_1) \sqcup \oRoot (r)) \cap B^*$. 
Then 
$B^*_1 = \oRooto ({q}_1) \sqcup \Root (V)$
with some $V \subset \Sub (r)$.
We are to show that 
there exists a $C^*_1 \in {\mathcal B}(N)$ satisfying
\begin{equation}
|C^*_1| =|B^*_1|, \qquad
\sum_{i \in C^*_1}A_i = \sum_{i \in B^*_1}A_i,
\end{equation}
and can be written as
$C^*_1 = \Root (V')$
with some $V' \subset \Sub (q_1) \sqcup \oSub (r)\subset \Trees (\Phi_A)$.

\begin{figure}[hbtp]
\centering
\scalebox{0.5}[0.5]{
\includegraphics[clip]{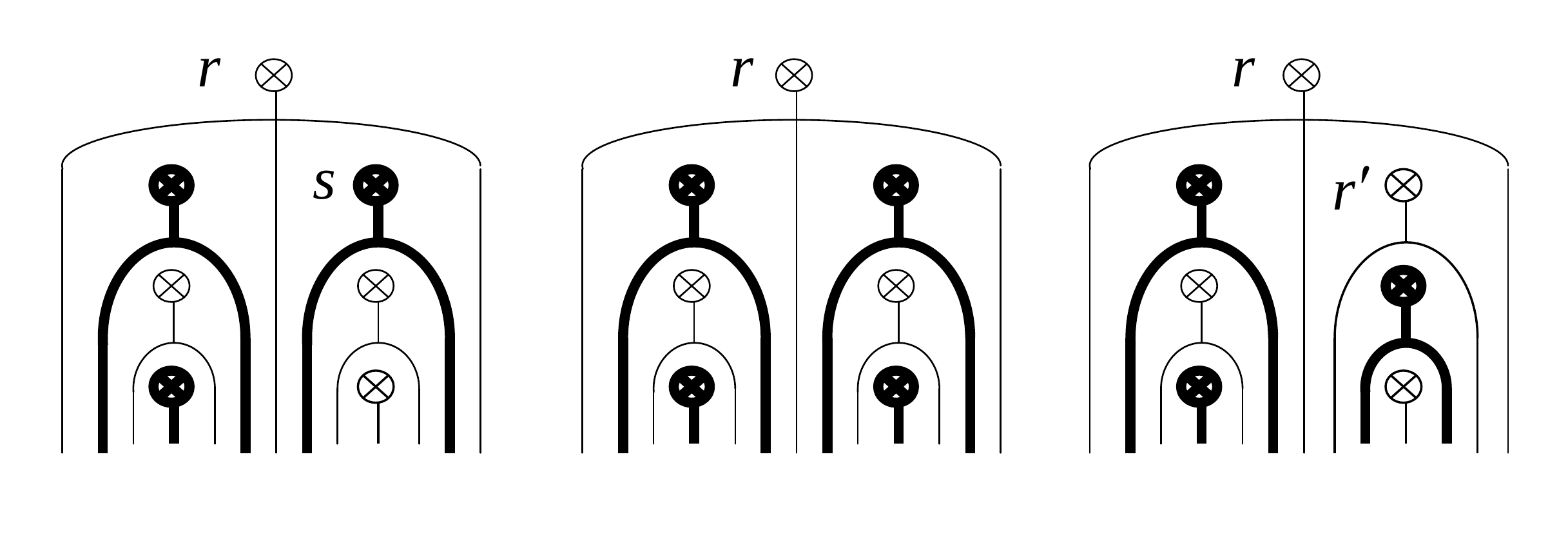}}
\caption{(Left) case (\ref{case1x}); (Middle) case \ref{case2ax}; (Right) case \ref{case2bx}}
\label{figure:8}
\end{figure}

\begin{enumerate}
\item \label{case1x}
Suppose there exists $s \in \MSub (r)$ such that both
$s \in V$ and
$\oSubo (s) \cap V \ne \oSubo (s) $ are satisfied.
See Figure \ref{figure:8} (Left) for an example.
Then $V_s = V \cap \oSub (s)$ is not closely packed with respect to $s$.
By Lemma \ref{lem:CPL} one finds that this case cannot happen because otherwise \eqref{eq:nov27_1} is not satisfied.
\item
Suppose otherwise, i.~e.~for any $s \in \MSub (r)$
either $s \notin V$ or $\oSubo (s) \subset V$ is
satisfied.

\begin{enumerate}
\item \label{case2ax}
Suppose $\oSubo (s) \subset V$ is satisfied for any
$s \in \MSub (r)$.
See Figure \ref{figure:8} (Middle).
This implies that
$\oSube (r) = V$,
hence
$B^*_1 = \oRooto ({q}_1) \sqcup \oRoote (r)$.
Let
$C^*_1 = \oRoote ({q}_1) \sqcup \oRooto (r)$.
Then $|C^*_1| =|B^*_1|$ and
by Lemmas \ref{lem:nov22_2}, 
\ref{lem:nov22_3} we have
\begin{equation}\label{eq:dec2_1}
\sum_{i \in C^*_1}A_i - \sum_{i \in B^*_1}A_i
\leq {\rm Ht} (r) - {\rm Ht} (P) = 0.
\end{equation}
The inequality case
is excluded because otherwise \eqref{eq:nov27_1} is not satisfied.

\item \label{case2bx}
Suppose otherwise, i.~e.~ there exists $r' \in \MSub (r)$ such that $r' \notin V$.
See Figure \ref{figure:8} (Right).
Replace $r$ by $r'$ and repeat the above arguments.
Since the order of the nesting is finite, this case \ref{case2bx}
cannot repeat endlessly, and we will eventually arrive at
case \eqref{case1x} or \ref{case2ax}. 
The case \eqref{case1x} has already been excluded.
The case \ref{case2ax} can be also excluded because
now we have ${\rm Ht} (r') - {\rm Ht} (P) < 0$ instead of the right equality of \eqref{eq:dec2_1}.
Thus one finds neither case can happen.
\end{enumerate}
\end{enumerate}

To summarize, the only possible case is 
 \ref{case2ax},
under the condition that the equality in
\eqref{eq:dec2_1} holds.
Then by changing $B^*$ by $(B^* \setminus B^*_1)\sqcup C^*_1$ one can reduce the number of adopted subtrees under $P$ by one
without changing the other conditions.
The proof is completed.

\qed

\section{A continuous analogue of Kerov-Kirillov-Reshetikhin bijection}\label{sec:4}
\subsection{A map from highest weight paths to rigged configurations}\label{sec:4_1}
The Kerov-Kirillov-Reshetkhin (KKR) bijection is a bijection between the set of tensor products of crystals \cite{Ka1} and the set of a combinatorial objects known as rigged configurations.
In this section we consider a continuous analogue of KKR map in $\mathfrak{sl}_2$ case
to explain the backgrounds of our algorithm in \S \ref{sec:2_4}.

Given $L >0$,
let $\mathcal{P}_+ = \coprod_{N=1}^\infty \mathcal{P}_{+,N}$ where
\begin{equation}\label{eq:jul11_1x}
\mathcal{P}_{+,N} = \left\{  (x_1,\dots,x_{2N}) \in (\mathbb{R}_{>0})^{2N} \bigg\vert \sum_{i=1}^{2N} x_i \leq L, 
\sum_{i=1}^k (x_{2i-1}-x_{2i}) 
\geq 0 \, \mbox{for} \, 1 \leq k \leq N 
\right\},
\end{equation}
and 
$\mathcal{M} = \coprod_{s=1}^\infty \mathcal{M}_s$ where
\begin{equation}\label{eq:jul11_2x}
\mathcal{M}_s = \left\{
\lambda = (l_j,m_j)_{1 \leq j \leq s} \in
 (\mathbb{R}_{>0} \times \mathbb{Z}_{>0} )^s \bigg\vert 
l_1 > \dots > l_s, \sum_{i=1}^s l_i m_i \leq L/2 \right\}.
\end{equation}
Each element of the set $\mathcal{M}_s$ is depicted as a Young diagram with area $\leq L/2$.

For $\lambda \in \mathcal{M}_s$ define its $j$th vacancy number $p_j(\lambda)$ as
\begin{equation}\label{eq:dec24_1}
p_j(\lambda)= L - 2 \sum_{k=1}^s \min (l_j,l_k) m_k.
\end{equation}
Also we define the set of quantum numbers or riggings ${\rm Rig} (\lambda)$ associated with $\lambda$ as
\begin{equation}\label{eq:dec23_1}
{\rm Rig} (\lambda) =
\left\{ (J_i^{(j)})_{1 \leq i \leq m_j, 1 \leq j \leq s} 
\in \R^{m_1+ \dots + m_s} 
\bigg\vert
0 \leq J_1^{(j)} \leq \dots \leq J_{m_j}^{(j)} \leq p_j(\lambda) \, \mbox{for} \, 1 \leq j \leq s
\right\}.
\end{equation}
Let $\Rig = \coprod_{\lambda \in \mathcal{M}} \Rig (\lambda)$.

Given $L>0$ we define a pair of maps
$\phi_1 : \mathcal{P}_+ \rightarrow \mathcal{M}$ and
$\phi_2 : \mathcal{P}_+ \rightarrow \Rig$, such that
$\phi = (\phi_1, \phi_2)$ gives a bijection $\phi :\mathcal{P}_+ \rightarrow \bigsqcup_{\lambda \in \mathcal{M}} \{ \lambda \} \times \Rig (\lambda)$.

\subsection{The map $\phi_1$} \label{sec:4_2}
In order to adjust the values of the quantum numbers to
conventional ones,
we slightly modify the algorithm 
in \S \ref{sec:2_4}
by replacing item \eqref{item:dec18_3} there by the following:
 
\begin{itemize}
\item
Suppose the last
sequence of $n_{k^{(i)}}^{(i)}$ zeros 
is at the right end 
after a positive neighbor $a$ as
$\ldots,a,0,\ldots,0$.
Remove these zeros.
If $n_{k^{(i)}}^{(i)}$ is odd then also remove 
$a$.
\end{itemize}
We define the map $\phi_1 : \mathcal{P}_+ \rightarrow \mathcal{M}$ by the algorithm 
in \S \ref{sec:2_4} with this modification.
Given the $A$-variables satisfying the highest weight condition, the Young diagram 
$\tilde{\lambda}$
constructed by the algorithm in \S \ref{sec:2_4} does not change under this modification.
To be clearer we call the algorithm in \S \ref{sec:2_4} algorithm-$I$, and the one in this section algorithm-$II$.
Then we have:
\begin{proposition}
The Young diagram constructed by
algorithm-$I$ is equal to the one by algorithm-$II$.  
\end{proposition}
\proof
To distinguish cases, we denote by $x^{(i)}(X), y^{(i)}(X)$ 
the sequences $x^{(i)}, y^{(i)}$ constructed by algorithm-$X(=\mbox{$I$ or $II$})$.
By induction on $i$,
it is easy to see that both algorithms preserve the highest weight condition, and that the lengths of the sequences are the same in both algorithms.
It is also easy to see that  $x^{(i)}_1(I) \geq x^{(i)}_1(II),  y^{(i)}_1(I) \geq y^{(i)}_1(II)$,
and  $x^{(i)}_j(I) = x^{(i)}_j(II),  y^{(i)}_j(I)= y^{(i)}_j(II)$ for $j > 1$.
This implies that  the set of numbers 
$\{ (\mu^{(i)},\nu^{(i)}) \}_{1 \leq i \leq s}$
constructed by both algorithms are the same.
\qed

By this fact and Theorem \ref{th:main}, we see that $\phi_1$ is a map that yields the conserved quantities
of trop p-Toda.

\begin{example}\label{ex:dec22_2}
For the same state in Example \ref{ex:dec22_1} we have:
\begin{align*}
x^{(1)} &= (4,3,3,2,2,1,1,1),\\
y^{(1)} &= (3,2,2,1,1,0,0,0),\\
x^{(2)} &= (3,2,2,1),\\
y^{(2)} &= (2,1,1,0),\\
x^{(3)} &= (2,1),\\
y^{(3)} &= (1,0),
\end{align*}
and $\mu^{(1)}=\mu^{(2)}=\mu^{(3)}=1,
\nu^{(1)}=2,\nu^{(2)}=\nu^{(3)}=1$.
\end{example}
The image of $\mathcal{P}_+$ by the map $\phi_1$ is indeed in $\mathcal{M}$.
\begin{lemma}\label{lem:jul11_3x}
$\phi_1 ( \mathcal{P}_+) \subset \mathcal{M}$.
\end{lemma}
\proof
It suffices to show that the condition $\sum_{i=1}^{2N} x_i \leq L$ in 
\eqref{eq:jul11_1x} leads to the condition $\sum_{i=1}^s l_i m_i \leq L/2$ in \eqref{eq:jul11_2x}.
It is done by using the relations $y^{(i)}_j = x^{(i)}_j - \mu^{(i)}$ and 
$\sum_{j=1}^{2 N^{(i+1)}}x^{(i+1)}_j \leq \sum_{j=1}^{2 N^{(i)}}y^{(i)}_j$ as
$L \geq \sum_{j=1}^{2 N^{(1)}}x^{(1)}_j = 2 N^{(1)} \mu^{(1)} +  
\sum_{j=1}^{2 N^{(1)}}y^{(1)}_j  \geq 2 \sum_{i=1}^s N^{(i)} \mu^{(i)} + 
\sum_{j=1}^{2 N^{(s)}}y^{(s)}_j \geq 2 \sum_{i=1}^s N^{(i)} \mu^{(i)}
= 2 \sum_{i=1}^s l_i m_i$.
\qed

\subsection{The map $\phi_2$}\label{sec:4_3}
Consider the algorithm
in \S \ref{sec:2_4} with the modification in \S \ref{sec:4_2}.
With the $i$th block 
$(\mu^{(i)},\nu^{(i)}) $ of the Young diagram $\tilde{\lambda}$
we associate $\nu^{(i)}$ non-negative real numbers called quantum numbers.
Recall that
in the array of non-negative real numbers
$y^{(i)}_1,\ldots,y^{(i)}_{2N^{(i)}}$ we have $k^{(i)}$ sequences of zeros, and the $j$th sequence has $n_j^{(i)}$ zeros.
Denote by $I_j^{(i)}$ the position of the first element of the $j$th sequence.
Let $r_j^{(i)} = \sum_{l=1}^{I_j^{(i)}-1}y_l^{(i)}$.
Then the $\nu^{(i)}$ quantum numbers for the $i$th block are defined as
\begin{equation}\label{eq:dec6_1}
\underbrace{r_1^{(i)} \ldots r_1^{(i)}}_{ \sigma_1^{(i)}}
\underbrace{r_2^{(i)} \ldots r_2^{(i)}}_{ \sigma_2^{(i)}}
\ldots
\underbrace{r_{k^{(i)}}^{(i)} \ldots r_{k^{(i)}}^{(i)}}_{\sigma_{k^{(i)}}^{(i)}},
\end{equation}
where $\sigma^{(i)}_j = \lceil n_{j}^{(i)}/2 \rceil$.
Note that $r_j^{(i)} < r_k^{(i)}$ for $j<k$.
Given any positive real number $L$ satisfying $\sum_{j=1}^{2N} x_i \leq L$,
let $\lambda^{(j)} = \sum_{k=1}^j \mu^{(k)}$
and
\begin{equation}\label{eq:jul14_2x}
q_j (\lambda) = L - 2 \sum_{k=1}^s 
\min (\lambda^{(j)},\lambda^{(k)})\nu^{(k)},
\end{equation}
for $1 \leq j \leq s$.
By Theorem \ref{th:main}, we have
$\lambda^{(s+1-j)} = l_j, \nu^{(s+1-j)} = m_j$,
and $q_{s+1-j}(\lambda) = p_j (\lambda)$.

By the following lemma,
one sees that
the quantum numbers for the $i$th block
obey the condition $0 \leq r_j^{(i)} \leq q_i (\lambda)$.
This enables us to define the map
$\phi_2 : \mathcal{P}_+ \rightarrow \Rig$
by the procedure described above and
by identifying 
the $m_{s+1-i} = \nu^{(i)}$ quantum numbers 
in \eqref{eq:dec6_1} with
$J_1^{(s+1-i)},\ldots,J_{m_{s+1-i}}^{(s+1-i)}$
for $1 \leq i \leq s$ in \eqref{eq:dec23_1}.
\begin{lemma}\label{lem:jul14_2x}
$\phi_2 (\mathcal{P}_+ ) \subset \Rig $.
\end{lemma}
\proof
By definition,
the relation $0 \leq r_j^{(i)} $ holds trivially.
Let $L_0 = L - \sum_{j=1}^{2N} x_i$.
It suffices to show that the relation
$\sum_{j=1}^{2N^{(i)}} y_j^{(i)} \leq q_i (\lambda)- L_0$
is satisfied for $1 \leq i \leq s$ by induction on $i$, since then we have
$r_j^{(i)} \leq \sum_{j=1}^{2N^{(i)}} y_j^{(i)} \leq  q_i (\lambda)$.
It is done by using the relations in
the proof of Lemma \ref{lem:jul11_3x} as well as
the relation
 $q_j (\lambda) = q_{j-1}(\lambda)-2 \mu^{(j)}N^{(j)}$ where $q_{0}(\lambda) = L$.
\qed

\begin{example}
For the $A$-variables in Example \ref{ex:nov17_1} 
the quantum numbers for the first block 
of the Young diagram are
$3 \sqrt{2}-1,3 \sqrt{2}-1,4 \sqrt{2}-2$,
those for the second are $2,2$, and that for the third is
$3 \sqrt{2}-4$.
\end{example}
\subsection{The inverse map}\label{sec:4_4}
Having obtained
the pair $\phi = (\phi_1, \phi_2)$ that gives a map $\phi :\mathcal{P}_+ \rightarrow \bigsqcup_{\lambda \in \mathcal{M}} \{ \lambda \} \times \Rig (\lambda)$, now we consider its inverse map $\phi^{-1}$.
This is done by using the inverse of the algorithm in \S \ref{sec:2_4} with the modification in \S \ref{sec:4_2}.
Let $\lambda = (l_j,m_j)_{1 \leq j \leq s} \in \mathcal{M}_s$
and $J = (J_i^{(j)})_{1 \leq i \leq m_j, 1 \leq j \leq s} 
\in {\rm Rig} (\lambda)$.
By the correspondence in the previous subsection, we regard $\lambda$ as
$\lambda = \{ (\mu^{(i)},\nu^{(i)}) \}_{1 \leq i \leq s}$,
and $J$ as
$J= \{ (r^{(i)}_j)^{\sigma^{(i)}_j} \}_{1 \leq j \leq k^{(i)}, 1 \leq i \leq s} \in  \Rig (\lambda)$
where $\sigma^{(i)}_j$ is the multiplicity of $r^{(i)}_j$. 
The quantum numbers obey the condition $0 \leq r^{(i)}_1 < \dots < r^{(i)}_{k^{(i)}} \leq q_i(\lambda)$
where $q_i(\lambda)$ is the vacancy number defined by \eqref{eq:jul14_2x}.
Given $(\lambda, J)$, its image $ x^{(1)}=(x^{(1)}_1,\dots, x^{(1)}_{2 N^{(1)}})$
under the map $\phi^{-1}$ is given by such a step-by-step construction as
$y^{(s)} \rightarrow x^{(s)} \rightarrow y^{(s-1)} \rightarrow x^{(s-1)} \rightarrow 
\dots \rightarrow y^{(1)} \rightarrow x^{(1)}$.
The first step goes as follows.
Let
\begin{equation*}
\underbrace{r_1^{(s)} \ldots r_1^{(s)}}_{ \sigma_1^{(s)}}
\underbrace{r_2^{(s)} \ldots r_2^{(s)}}_{ \sigma_2^{(s)}}
\ldots
\underbrace{r_{k^{(s)}}^{(s)} \ldots r_{k^{(s)}}^{(s)}}_{\sigma_{k^{(s)}}^{(s)}}
\end{equation*}
be the $\nu^{(s)}$ quantum numbers for the $s$-th (top) block.
We define
a sequence of
non-negative real numbers
$y^{(s)}_1,\ldots,y^{(s)}_{2N^{(s)}}$
as
\begin{displaymath}
r^{(s)}_{1}, \underbrace{0,\dots,0}_{ 2 \sigma_{1}^{(s)}-1},
r^{(s)}_{2} - r^{(s)}_{1}, \underbrace{0,\dots,0}_{ 2 \sigma_{2}^{(s)}-1},
\ldots ,
r^{(s)}_{k^{(s)}} - r^{(s)}_{k^{(s)}-1},
\underbrace{0,\dots,0}_{ 2 \sigma_{k^{(s)}}^{(s)}-1}.
\end{displaymath}
Then we define 
a sequence of
positive real numbers
$x^{(s)}_1,\ldots,x^{(s)}_{2N^{(s)}}$
by $x^{(s)}_j = y^{(s)}_j + \mu^{(s)}$.

The subsequent steps go as follows.
Given
$x^{(i+1)}_1,\ldots,x^{(i+1)}_{2N^{(i+1)}}$ 
and the quantum numbers 
in \eqref{eq:dec6_1}, we define
a sequence of
non-negative real numbers
$y^{(i)}_1,\ldots,y^{(i)}_{2N^{(i)}}$
by the following way.
Let $w_k^{(i+1)} = \sum_{j=1}^k x_j^{(i+1)}$ for
$1 \leq k \leq 2N^{(i+1)}$ and
$w_0^{(i+1)}=0$.
For each $r^{(i)}_j$
either $w_k^{(i+1)} \leq r^{(i)}_j < w_{k+1}^{(i+1)}$
for some $0 \leq k \leq 2N^{(i+1)}-1$
or $r^{(i)}_j \geq w_{2N^{(i+1)}}^{(i+1)}$ is satisfied.
Roughly speaking, we split $x_{k+1}^{(i+1)}$ and insert some zeros
in the former case,
while in the the latter case we append $r^{(i)}_j - w_{2N^{(i+1)}}^{(i+1)}$
and some zeros at the end of the sequence.
To be more precise, let us consider the case
$w_k^{(i+1)} \leq r^{(i)}_j <  r^{(i)}_{j+1} < w_{k+1}^{(i+1)}$
and the case $r^{(i)}_{k^{(i)}} > r^{(i)}_{k^{(i)}-1} \geq w_{2 N^{(i+1)}}^{(i+1)}$ as examples,
where we assumed no other $r^{(i)}_l$'s exist in the (half-)intervals determined by $w_k^{(i+1)}$'s.
In the former case we replace $x_{k+1}^{(i+1)} = w_{k+1}^{(i+1)} - w_{k}^{(i+1)}$ by
\begin{displaymath}
r^{(i)}_j - w_k^{(i+1)}, \underbrace{0,\dots,0}_{ 2 \sigma_j^{(i)}-1},r^{(i)}_{j+1} - r^{(i)}_{j},
\underbrace{0,\dots,0}_{ 2 \sigma_{j+1}^{(i)}-1},
w_{k+1}^{(i+1)}-r^{(i)}_{j+1}.
\end{displaymath}
In the latter case we add the following sequence
after $x_{2 N^{(i+1)}}^{(i+1)}$:
\begin{displaymath}
r^{(i)}_{k^{(i)}-1} - w_{2 N^{(i+1)}}^{(i+1)}, \underbrace{0,\dots,0}_{ 2 \sigma_{k^{(i)}-1}^{(i)}-1},r^{(i)}_{k^{(i)}} - r^{(i)}_{k^{(i)}-1},
\underbrace{0,\dots,0}_{ 2 \sigma_{k^{(i)}}^{(i)}-1}.
\end{displaymath}
It is easy to generalize these procedures
for the cases where any number of different
values of the quantum numbers exist in the (half-)intervals determined by $w_k^{(i+1)}$'s
.
Then we define 
a sequence of
positive real numbers
$x^{(i)}_1,\ldots,x^{(i)}_{2N^{(i)}}$
by $x^{(i)}_j = y^{(i)}_j + \mu^{(i)}$.

Given
$\lambda = \{ (\mu^{(i)},\nu^{(i)}) \}_{1 \leq i \leq s} \in \mathcal{M}_s$
and
$J= \{ (r^{(i)}_j)^{\sigma^{(i)}_j} \}_{1 \leq j \leq k^{(i)}, 1 \leq i \leq s} \in  \Rig (\lambda)$
, we define the map
$\phi^{-1}$ by $\phi^{-1}(\lambda, J) = (x^{(1)}_1,\dots, x^{(1)}_{2 N^{(1)}})$. 
By construction, we see that it is indeed the inverse
of the map $\phi$.
Moreover we have the following:
\begin{lemma}\label{lem:jul14_3x}
$\phi^{-1} (\bigsqcup_{\lambda \in \mathcal{M}} \{ \lambda \} \times \Rig (\lambda)) 
 \subset \mathcal{P}_+$.
\end{lemma}
\proof
From Lemma \ref{lem:jul14_1x} we see that the above algorithm for $\phi^{-1}$
preserves the highest weight condition.
Hence it suffices to show that
$w^{(1)}_{2 N^{(1)}} := \sum_{j=1}^{2 N^{(1)}}x^{(1)}_j \leq L$.
To begin with, we prove $\sum_{j=1}^{2N^{(i)}} y_j^{(i)} \leq q_i (\lambda)$
for $1 \leq i \leq s$.
For $i=s$, it is satisfied as $\sum_{j=1}^{2N^{(s)}} y_j^{(s)} = r^{(s)}_{k^{(s)}} \leq q_s (\lambda)$.
Suppose $\sum_{j=1}^{2N^{(i+1)}} y_j^{(i+1)} \leq q_{i+1} (\lambda)$ for some $i < s$.
Then we have
$w^{(i+1)}_{2 N^{(i+1)}}
 = \sum_{j=1}^{2N^{(i+1)}} y_j^{(i+1)} + 2 N^{(i+1)} \mu^{(i+1)}\leq q_{i+1} (\lambda) + 2 N^{(i+1)} \mu^{(i+1)} = q_i(\lambda)$, and hence
$\sum_{j=1}^{2N^{(i)}} y_j^{(i)} =\max \{w^{(i+1)}_{2 N^{(i+1)}}, r^{(i)}_{k^{(i)}} \} \leq q_i (\lambda)$.
Thus by descending induction on $i$ this inequality holds for any  $1 \leq i \leq s$.
Now we obtain the desired result as $w^{(1)}_{2 N^{(1)}}
= \sum_{j=1}^{2N^{(1)}} y_j^{(1)} + 2 N^{(1)} \mu^{(1)}\leq q_{1} (\lambda) + 2 N^{(1)} \mu^{(1)} =q_0(\lambda) = L$.
\qed

By Lemmas \ref{lem:jul11_3x}, \ref{lem:jul14_2x} and \ref{lem:jul14_3x}
we obtain the following result.
\begin{theorem}
The map $\phi : \mathcal{P}_+ \rightarrow \bigsqcup_{\lambda \in \mathcal{M}} \{ \lambda \} \times \Rig (\lambda)$ is a bijection.
\end{theorem}
So far we do not know whether one can regard this bijection as an isomorphism, i.~e.~we do not know what kind of mathematical structures are preserved under this bijection.

\section{Concluding remarks}\label{sec:5}
In this paper we elucidated combinatorial aspects of the conserved quantities of general tropical periodic Toda lattice
beyond the generic condition.
Let us summarize what we have done.
The evolution equation of this dynamical system
was given by \eqref{eq:july2_7},
and the conserved quantities were written
as in \eqref{eq:july2_8}.
We proved that the conserved quantities are related by a weak convexity condition (Theorem \ref{th:jun10_1}), which enables us to write the set of conserved quantities as a Young diagram in Figure \ref{figure:2}.
After introducing an algorithm related to the Kelov-Kirillov-Reshetikhin (KKR) bijection to construct another Young diagram in Figure \ref{figure:3}, we presented our main result (Theorem \ref{th:main}) saying the identification of these Young diagrams.
We gave a detailed proof of this theorem 
and a discussion on the KKR bijection in the subsequent sections.

The idea of our proof is based on \cite{IT07}, but not a straightforward extension.
From our interpretation, there are several ambiguous
and/or incorrect descriptions in Appendix A.~1 of \cite{IT07}.
To make our proof mathematically rigorous,
we devise our own tools.
The followings are two of them.
(i) In \S \ref{sec:3_1} we devised our original rule to draw lines in a graph when more than two consecutive $\times$'s appear in a given level.
In the 10-elimination algorithm for pBBS this corresponds to 
simultaneous disappearance of more than two consecutive blocks, for which the rule to draw lines was ambiguous in \cite
{IT07}.
(ii) We formulated our original lemmas in \S \ref{sec:3_4} and
\S \ref{sec:3_5}.
Here we explain the latter one.
From our interpretation, equation (A. 13) of \cite{IT07} claims that
{\em any} $B^*$ in \eqref{eq:nov27_1} {\em must} be realized by a forest, on which
their proof substantially depend.
But as we have shown in Example \ref{ex:jan11_1} this claim is not true.
The correct statement is that {\em at least one} 
$B^*$ in \eqref{eq:nov27_1} {\em can} be realized by a forest,
as we proved in  \S \ref{sec:3_5}.

In the original (discrete) KKR bijection  in $\mathfrak{sl}_2$ case \cite{KTT} the set $ \mathcal{P}_+ $ is a subset of
the tensor product of crystals
$B^{\otimes L}$ with $B = \{ 1,2 \}$ and its elements are expressed as
\begin{equation*}
\underbrace{1 \dots 1}_{x_1}\underbrace{2 \dots 2}_{x_2}\dots
\underbrace{1 \dots 1}_{x_{2N-1}}\underbrace{2 \dots 2}_{x_{2N}}
\underbrace{1 \dots 1}_{x_{2N+1}}
\end{equation*}
for some $N$
with the condition in \eqref{eq:jul11_1x}, where we omitted $\otimes$ symbols.
In our algorithm the integer $x_i$'s here have been replaced by
continuous variables taking their values in
real numbers.
Actually our algorithm is based on the algorithm in \cite{Takagi1}
which is a variation of the original algorithm.
In $\mathfrak{sl}_{n+1}$ case the set $B$ is replaced by $B=\{ 1,\dots, n+1 \}$
and the highest weight condition is adequately modified.
The algorithm of the KKR bijection for $\mathfrak{sl}_{n+1}$ case was given in \S 3.2 of \cite{IKT12}.
In this case no analogues of the above mentioned variation in  $\mathfrak{sl}_2$ case
has been developed yet.
Therefore a promising way to
construct a continuous analogue of the KKR bijection for $\mathfrak{sl}_{n+1}$ is
to develop such a variation first.
Then the remaining task will be rather straightforward.

Finally we would like to explain the difference of the arguments
between \cite{KTT} and the present work.
In \cite{KTT} the conservation of the Young diagram 
under the time evolution of 
pBBS was shown by the following way.
For any positive integer $l$ and any state $p$,
we introduced a time evolution $T_l(p)$ and an energy $E_l(p)$ (Proposition 2.1) by using the crystal theory and its energy function.
Then the conservation of the energy $E_l(T_k(p)) = E_l (p)$ 
and the commutativity of the time evolutions $T_l T_k (p) = T_k T_l (p)$
were shown (Theorem 2.2).
Finally we proved that the data carried by
the whole set of the values of the energy $E_l(p) \, (l=1,2,\dots)$
was equivalent to the Young diagram constructed by KKR map (Proposition 3.4).
In the present work the author did not try to generalize this argument
to trop p-Toda case, because the crystal theory and its energy function have not been developed in this case.
Therefore no counterpart of the above construction of $E_l(p)$ can be allocated in the present paper.
However, the author thinks that one can
generalize the above argument to trop p-Toda case
because the arguments in \cite{T2012}
to construct a commuting family of time evolutions
may be used to develop an analogue of the
energy function in this case.
We hope to report any progress on this subject in the near future. 


\vspace{5mm}
\noindent
{\it Acknowledgement}.
This work was supported by 
JSPS KAKENHI Grant Number  25400122.


\vspace{0.5cm}

\end{document}